\documentclass[draftclsnofoot,a4paper,12pt,onecolumn,doublespace]{IEEEtran}

\makeatletter

\def\ps@IEEEtitlepagestyle{%
  \def\@oddfoot{\mycopyrightnotice}%
  \def\@evenfoot{}%
}
\def\mycopyrightnotice{%
  {\footnotesize This work has been submitted to the IEEE for possible publication. Copyright may be
transferred without notice, after which this version may no longer be accessible. \hfill}
  \gdef\mycopyrightnotice{}
}

\def\ps@headings{%
\def\@oddhead{\mbox{}\scriptsize\rightmark \hfil \thepage}%
\def\@evenhead{\scriptsize\thepage \hfil \leftmark\mbox{}}%
\def\@oddfoot{}%
\def\@evenfoot{}}
\makeatother
\usepackage{caption}
\usepackage{subcaption}

\usepackage{graphicx,cite}
\usepackage{multirow}
\usepackage{epsfig}
\usepackage[cmex10]{amsmath}
\usepackage{array}
\usepackage{subfig}

\usepackage[algo2e]{algorithm2e}
\usepackage{algorithm}
\usepackage{algorithmic}

\usepackage{amssymb}
\usepackage{array}
\usepackage{fixltx2e}
\usepackage{amsfonts}

\begin{document}
\title{System Power Minimization to Access Non-Contiguous Spectrum in Cognitive Radio Networks}

\author{\IEEEauthorblockN{Muhammad Nazmul Islam\IEEEauthorrefmark{0}, 
Narayan B. Mandayam\IEEEauthorrefmark{1}, Sastry Kompella\IEEEauthorrefmark{2} 
and Ivan Seskar\IEEEauthorrefmark{1} \\
 }
 \IEEEauthorblockA{\IEEEauthorrefmark{0}
Qualcomm, Email: mislam@qti.qualcomm.com, 
\IEEEauthorrefmark{1} 
WINLAB, Rutgers University, 
Email: \{narayan,seskar\}@winlab.rutgers.edu,
\IEEEauthorrefmark{2} 
Naval Research Laboratory, Email: sk@ieee.org}}

\maketitle

\begin{abstract}


Wireless transmission using non-contiguous 
chunks of spectrum is becoming increasingly important due to a variety of 
scenarios such as: secondary users avoiding incumbent users in TV white space; 
anticipated spectrum sharing between commercial and military systems; 
and spectrum sharing among uncoordinated interferers in unlicensed bands. 
Multi-Channel Multi-Radio (MC-MR) platforms and 
Non-Contiguous Orthogonal Frequency Division Multiple Access (NC-OFDMA) technology
are the two commercially viable transmission choices to access these non-contiguous spectrum chunks. 
Fixed MC-MRs do not scale with increasing
number of non-contiguous spectrum chunks due to their fixed
set of supporting radio front ends. NC-OFDMA
allows nodes to access these non-contiguous spectrum chunks
and put null sub-carriers in the remaining chunks.
However, nulling sub-carriers increases the sampling rate (spectrum span) 
which, in turn, increases the power consumption of radio front ends.
Our work characterizes this trade-off from a cross-layer perspective, 
specifically by showing how the slope of ADC/DAC's power consumption versus sampling rate curve 
influences scheduling decisions in a multi-hop network. 
Specifically, we provide a branch and bound algorithm based 
mixed integer linear programming solution
that performs joint power control, spectrum span selection, scheduling and routing in order to minimize the system power of multi-hop NC-OFDMA networks.
We also provide a low complexity $(O(E^2 M^2))$ greedy algorithm 
where $M$ and $E$ denote the number of channels and links respectively. 
Numerical simulations suggest that our approach reduces system power by $30\%$ over
classical transmit power minimization based cross-layer algorithms.

\end{abstract}

\newpage

\vspace{-2mm}


\section{Introduction}   \label{sec:Intro}

Demand for wireless services is becoming much greater than the currently available spectrum~\cite{NazmulCISSNew}. 
FCC has already opened up 300 MHz in TV bands~\cite{802.22} and plans to open up an additional 500 MHz by 2020~\cite{FCCSmallCell} to meet this demand. Any radio can use these channels if it abides by FCC specifications~\cite{FCCSmallCell}. If uncoordinated networks (e.g. different broadband wireless service providers) use these channels, they will adjust spectrum usage according to their individual traffic demands. Available spectrum will become partitioned into a set of non-contiguous segments. For some bands, like white 
space~\cite{802.22}, available spectrum itself is non-contiguous.


\begin{figure}[t]
\centering
\includegraphics[scale=0.45]{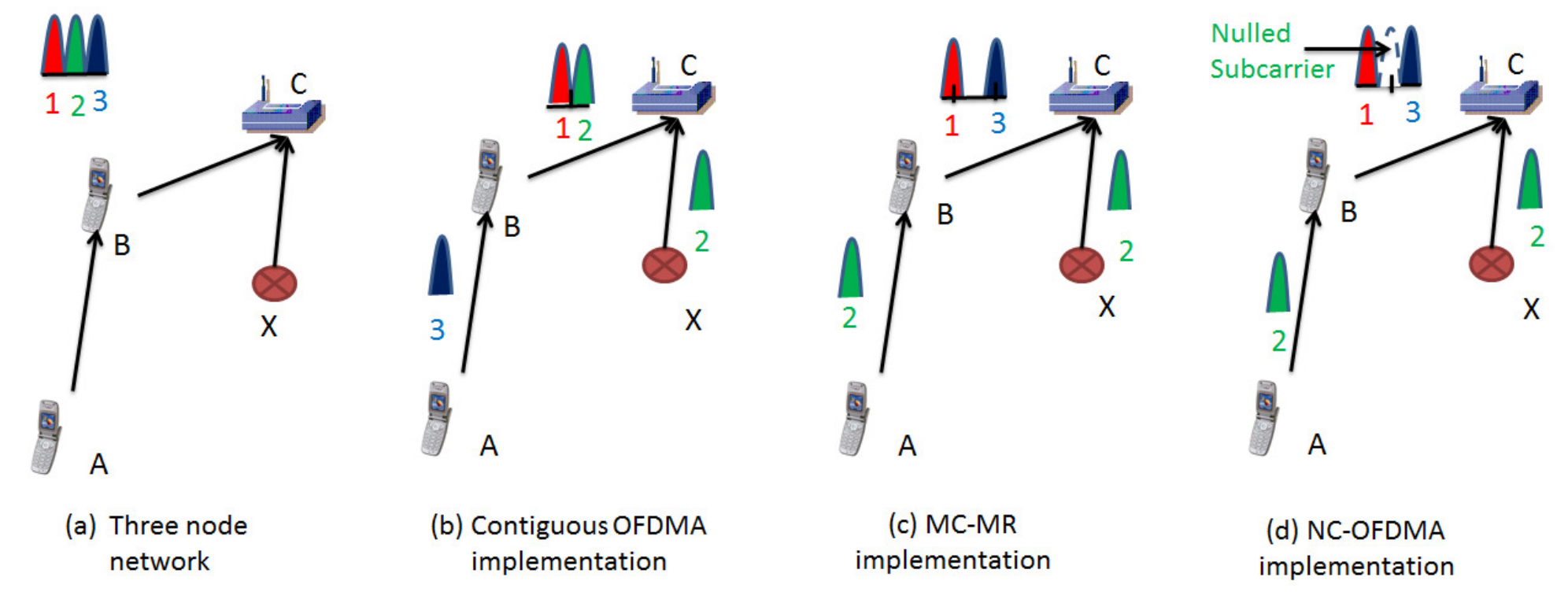}
\caption{Motivation of Non-contiguous Spectrum Access}
\label{fig:NCOFDMA_Motivation}
\end{figure}

Multi-Channel Multi-Radio (MC-MR) technology allows nodes to simultaneously access multiple fragmented spectrum chunks~\cite{MC-MR1,MC-MR2}. 
However, the number of non-contiguous spectrum chunks that fixed MC-MR can access 
is limited by the number of available radio front ends; which is
often constrained due to the size and power 
limitations of the transmission device~\cite{MCMRLimitations}.
Software defined radio based Non-Contiguous Orthogonal Frequency Division Multiple Access (NC-OFDMA) technology allows nodes to transmit in these non-contiguous spectrum chunks with a single radio front end. Nodes can null interference-limited subcarriers and select better channels in NC-OFDMA enabled networks. Hence, NC-OFDMA has grabbed a lot of attention in resource allocation~\cite{Hou,Kompella,Hou2}, cooperative 
forwarding~\cite{NazmulWiOpt12,Baochun} and experimental research~\cite{UCSB2,UCSB3}. However, nulling unwanted channels 
increases the spectrum span and the sampling rate of nodes.

Fig.~\ref{fig:NCOFDMA_Motivation} illustrates the benefits and inherent challenges of NC-OFDMA using the example of a two-hop network.
Fig. 1(a) shows a two-hop network where
node A transmits to node C via node B; node B relays node A's
data and also transmits its own data to node C. Channel 1, 2 and 3 are available channels.
Node X, an external and uncontrollable interferer, transmits in channel 2.
If we want to minimize the maximum rate between node A and B and allocate channels accordingly,
node B will require two channels and node A will require one channel.
There are three possible techniques/configurations to allocate the three available
channels as shown in Figures 1(b) through 1(d). Fig. 1(b) shows contiguous OFDMA that suffers from interference in link BC at channel 2.
Fig. 1(c) shows MC-MR that requires two radio front ends to capture channel $1$ and $3$ in link BC.
Fig. 1(d) shows NC-OFDMA that uses only one radio front end, transmits at channel $1$ and $3$ and nulls channel $2$.
However, due to the nulling of channel $2$, NC-OFDMA spans three channels, instead of two.

It is well known that the circuit power consumption of ADC and DAC increase 
linearly and exponentially with sampling rate and the number of quantization bits respectively~\cite{Goldsmith,ConverterPassion}. As software defined radios continue to improve their higher quantization resolution, the ADC and DAC that are used in the radio circuits will dominate the amount of power consumption. A comparison between Table~\ref{tab:ADCPower} and Table~\ref{tab:802.22Rules} shows that the power consumption of some commercial ADCs is more than 10 dB higher than the maximum allowed transmission power for portable devices 
in the 802.22 standard. On the one side, NC-OFDMA reduces transmission power by selecting channels with better link gains, while on the other side, this increased spectrum span increases circuit power consumption of the transceiver. 
In this paper, we investigate this trade-off between transmission power reduction and circuit power increase in the context of 
cross-layer optimization of NC-OFDMA based wireless networks. 

\begin{table}
\begin{center}
\begin{tabular}{|l|l|l|l|} \hline
Device & Device & Max. Sampling & Power \\
Name & Type & Rate (MS/s) & Dissipation (mW) \\ \hline
AD 9777~\cite{USRP_DAC} & DAC & 150 & 1056  \\ \hline
ADS62P4~\cite{USRP_ADC} & ADC & 125 & 908 \\ \hline
ADC 9467B~\cite{AD9467} & ADC & 250 & 1333 \\ \hline
\end{tabular}
\end{center}
\caption{Maximum Sampling Rates and Power Dissipation of Different
ADC/DAC} \label{tab:ADCPower}
\end{table}

\begin{table}
\begin{center}
\begin{tabular}{|l|l|l|} \hline
Device  & Allowed  & Operating  \\ 
Type & Power (mW) & Frequency (MHz) \\ \hline
Fixed & 4000 & 54 - 698  \\ \hline
Portable & 100 & 512-698 \\ \hline
\end{tabular}
\end{center}
\caption{Maximum Allowed Power and Operating Frequencies in IEEE 802.22~\cite{Cyrus}} 
\label{tab:802.22Rules}
\end{table}

Specifically, we ask the following question in this work: \emph{How can
single front end radio based nodes of a multi-hop network
access non-contiguous spectrum chunks?} We investigate this question
from a system power perspective and find that \emph{scheduling a small subset of channels may outperform
traditional transmit power minimization based approaches -- which allocate power across all
`good' channels -- since it consumes less circuit power.}
Our algorithm selects scheduling variables based on the slope of
ADC and DAC power consumption versus sampling rate curves. We show two special
sub-cases of this finding. We find that if the curves are almost flat, our algorithm converges to the 
transmission power minimization based scheduling algorithms. 
If the curves are very steep, our
algorithm selects the channel with the highest link gain.
For commercial ADC \& DAC's whose slopes lie between these two extreme
cases, our algorithm selects channels to minimize the summation
of transmit and circuit powers of the network.

In general, we provide a branch and bound algorithm based mixed 
integer linear programming solution
that performs joint power control, spectrum span selection, 
scheduling \& routing and minimizes system power of multi-hop NC-OFDMA networks.
We also provide a greedy algorithm that runs in
$O(E^2M^2)$ time where $E$ and $M$ denote the number 
of links and channels respectively. 
\emph{To the best of our knowledge,
ours is the first work that shows how the slope of ADC/DAC's power 
consumption curve influences the scheduling decisions of a multi-hop network}.
Numerical simulations suggest that our approach reduces system power by $30$\% over
classical transmission power minimization based cross-layer algorithms.

\subsection{Related Work}

The authors of~\cite{MC-MR1,MC-MR2} characterized the capacity region
of an MC-MR based multi-hop network.
The authors of~\cite{Hou,Kompella} focused on software defined radio
based multi-hop networks and performed cross layer optimization
using a protocol and signal-to-interference-plus-noise-ratio model
respectively. 
None of these works considered circuit power and addressed how
non-contiguous spectrum access influences cross-layer decisions.

Consideration of system power has been gaining attention in energy
efficient wireless communications literature~\cite{CktPower2}.
Cui et. al. focused on system energy constrained modulation
optimization in~\cite{Goldsmith}. Sahai et. al. investigated
system power consumption -- especially decoder power consumption --
in ~\cite{Sahai}. Isheden and Fettweis assumed circuit power
to be a linear function of the data rate~\cite{CktPower1}.
All these works focused on single transceiver pair. Our approach
differs from these works in the following way:
in NC-OFDMA technology, ADC and DAC
consume power not only for used channels (i.e. transmitted data) but also
for nulled channels. Our work considers the power consumption
related to spectrum span and investigates the performance of NC-OFDMA based
multi-hop networks.

The impact of hardware constraints on the performance of
NC-OFDMA networks was previously raised in~\cite{Bundle,GuardBand}.
The authors of~\cite{Bundle} performed cross-layer resource allocation
when each node's maximum spectrum span is limited by its ADC/DAC.
The authors of~\cite{GuardBand} investigated how
the size of the guardband, required to reduce cross-band interference,
affects the performance of NC-OFDMA based distributed transceiver pairs.
Our work uses system power to investigate the performance of NC-OFDMA based
multi-hop networks.

Periodic non-uniform sampling (PNS) can recover
a sparsely located non-contiguous signal with sub-Nyquist sampling rate~\cite{PNS}
and potentially reduce the power consumption of the ADC circuits.
However, PNS requires Nyquist rate circuitry at track-and-hold stages. 
PNS also needs compensation for imperfect production of the time delay elements.
As a result, PNS is not widely used by industry to access non-contiguous spectrum~\cite{SubNyquistSummary}. 
Our work focuses on the system power minimization of an NC-OFDMA network that uses commercial ADCs and DACs with Nyquist sampling rate 
to access non-contiguous spectrum.

Time and frequency mismatch affect an NC-OFDMA network
more severely due to its use of a large number of nulled sub-carriers.
Several researchers have implemented different techniques
to reduce interference between unsynchronized NC-OFDMA nodes.
The authors of~\cite{UCSB3} have used adaptive multi-bank
stop-band filters to reduce interference of unwanted channels.
The authors of~\cite{UCSB2} have used wider guard bands
to reduce leakage power into neighboring channels. 
We do not focus on synchronization techniques and 
testbed implementation of NC-OFDMA nodes in our work. 
Interested readers are suggested to go through~\cite{UCSB1,UCSB2,UCSB3,UCSB4}
to understand the implementation and synchronization details of NC-OFDMA.

The remainder of the paper is organized as follows:
Section~\ref{sec:Model} presents system power
and  multi-hop network model. 
Section~\ref{sec:Solution} provides
a branch and bound based solution of the optimization problem.
We present theoretical insights of our algorithm
in Section~\ref{sec:Insights} and a low complexity 
algorithm in Section~\ref{sec:GreedyAlgo}.
After showing numerical results in section~\ref{sec:Simulation},
we conclude in section~\ref{sec:Conclusion}.

\section{System Model}  \label{sec:Model}

\subsection{System Power Model}

We assume that baseband signal processing techniques like multi-user detection and
iterative decoding are not employed in the system. In this context,
power consumption in the baseband is negligible compared with
that in the radio frequency (RF) circuitry~\cite{LowPowerBaseband}.
\begin{figure}[t]
\centering
\includegraphics[scale=0.35]{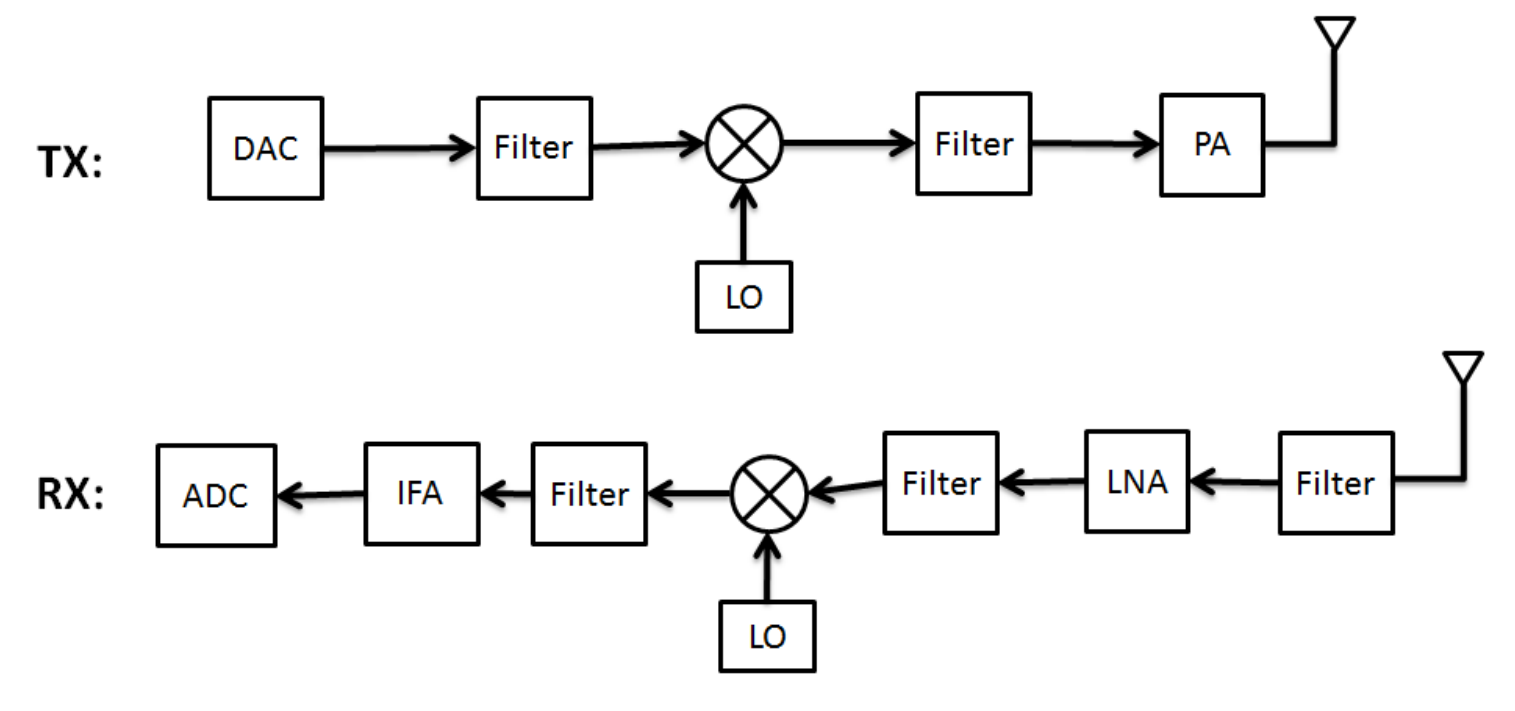}
\caption{Radio front end circuit blocks  (reproduced from ~\cite{Goldsmith})}
\centering
\label{fig:System_Block}
\end{figure}
Each radio node has two front ends: one for transmission and the other
for reception. Nodes are half-duplex, i.e., they can simultaneously transmit and
receive using these two front ends but not in the same channel.

Fig.~\ref{fig:System_Block} shows signal level
blocks in the transmitter and receiver. 
At the transmitter, the baseband digital signal goes through
DAC, filters, mixer (where it gets multiplied by the local oscillator (LO)) 
and programmable amplifier (PA) before reaching transmitter antenna. 
The received signal at the antenna goes through low noise amplifiers (LNA), filters,
mixer, intermediate frequency amplifier (IFA) and ADC to
reach the baseband circuit~\cite{Goldsmith}.

Typically, transceiver circuits work on a multi-mode basis. 
When there is a signal to transmit, all RF circuits work
in active mode; when there is no signal to transmit, all
RF circuits remain in sleep mode; circuits switch from sleep
to active mode through transient mode~\cite{Goldsmith}.
Here, we focus on system power minimization of all
RF circuits in the active mode. 
Let $p_t$, $p_r$ and $p$ denote
active mode power consumption of transmitter and receiver, and transmitter's emitted power at radio frequency respectively. Now,
\begin{equation}
p_t = \frac{PAPR}{\eta} p + p_{tc}, \, \, \, p_r  =  p_{rc} \label{eq:PowerRx}
\end{equation}
%
where $PAPR$ and $\eta$ denote the peak-to-average-power-ratio (PAPR) and
drain efficiency of the programmable amplifier. $\frac{PAPR}{\eta} p$
denote the total power consumption of programmable amplifier~\cite{Goldsmith}.
Also, $p_{tc}$ and $p_{rc}$ are the circuit power 
consumption of transmitter (excluding programmable amplifier's 
power consumption) and receiver, respectively.

The power consumption 
at almost all blocks of the radio front end, except ADC and DAC, does
not depend on sampling rate~\cite{Goldsmith,SystemLevelPower}. 
The power consumption of ADC
and DAC are affine functions of the sampling rate~\cite{ConverterPassion,Goldsmith}.
Denoting $k_{pa} = \frac{PAPR}{\eta}$,
system power consumption in the transmitter and receiver front end
become:
\begin{eqnarray}
p_t & = & \alpha_1 + \alpha_2 f_{st} + k_{pa} p \label{eq:SystemPowerTx}  \\
p_r & = & \beta_1 + \beta_2 f_{sr}     \label{eq:SystemPowerRx}
\end{eqnarray} 
In the above, $f_{st}$ and $f_{sr}$ are the sampling rates of the 
transmitter and receiver path. 
$\alpha_1$, $\alpha_2$, $\beta_1$ and $\beta_2$ are constants
that depend on the power consumption of different blocks. 
Appendix~\ref{sec:PowerConsumption} describes the power 
consumption of each block in details.
Table~\ref{tab:Notations} shows the list of notations that we have
used throughout the paper.

\begin{table}
\begin{center}
\begin{tabular}{|l|l|} \hline
Notation & Description \\  \hline
$\mathcal{N}$ & Set of nodes \\ \hline
$N$ & Total number of nodes \\ \hline
$\mathcal{E}$ & Set of edges \\ \hline
$E$ & Total number of edges \\ \hline
$\mathcal{L}$ & Set of sessions \\ \hline
$L$ & Total number of sessions \\ \hline
$(s(l), d(l))$ & Source and destination of session $l$ \\ \hline
$r(l)$ & Rate requirements of session $l$ \\ \hline
$W$ & Bandwidth of each channel \\ \hline
$N_0$ & Noise spectral density \\ \hline
$f_{ij}^m(l)$ & Flow on link $ij$ in channel $m$ for session $l$  \\ \hline
$c_{ij}^m$ & Capacity on link $ij$ in channel $m$  \\ \hline
$s_{ij}^m$ & Signal-to-noise ratio on link $ij$ in channel $m$ \\ \hline
$\mathcal{M}$ & Set of all available channels across all nodes \\ \hline
$\mathcal{M}_i$ & Set of available channels in node $i$ \\ \hline
$P_I$ & Interference threshold \\ \hline
$\mathcal{M}_{ij}$ & Set of available channels between node $i$ and $j$ \\ \hline
$M$ & Total number of available channels \\ \hline
$g_{ij}^m$ & Link gain of $ij$ in channel $m$ \\ \hline
$p_{ij,m}$ & Allocated power between $i$ and $j$ in channel $m$ \\ \hline
$x_{ij}^m$ & If link $ij$ uses channel $m$ \\ \hline
$x_i^{t,m}$ & If node $i$ uses channel $m$ for transmission \\ \hline
$x_i^{r,m}$ & If node $i$ uses channel $m$ for reception \\ \hline
$q_{t,i}$ & Spectrum span of the transmitter path of node $i$ \\ \hline
$q_{r,i}$ & Spectrum span of the receiver path of node $i$ \\ \hline
$f_{st,i}$ & Sampling rate of node $i$'s transmitter path \\ \hline
$f_{sr,i}$ & Sampling rate of node $i$'s receiver path \\ \hline
$f_{s,max}$ & Maximum allowed sampling rate of the nodes  \\ \hline
$P_{s,max}$ & Maximum allowed system power consumption \\ \hline
$A$ & An arbitrary large number \\ \hline
\end{tabular}
\end{center}
\caption{List of Notations} \label{tab:Notations}
\end{table}

\subsection{Multi-hop cross-layer model}

We consider a multi-hop network with a set of $\mathcal{N}$ cognitive 
radio nodes.   
Let $\mathcal{M}$ denote the set of all available channels.
Bandwidth of each channel is $W$.

\subsubsection{Power Control and Scheduling Constraints}

The number of available channels may vary in different (spatial)
parts of the network due to interference and other spectrum constraints
such as primary users and systems. Hence, we focus on frequency scheduling. 
Denote the binary scheduling decision $x_{ij}^m$ as follows:
\begin{equation}
  x_{ij}^m =\begin{cases}
    1, & \text{if node $i$ transmits to node $j$ using channel $m$}.\\
    0, & \text{otherwise}.
  \end{cases}
\end{equation} 
Due to self-interference, node $i$ can use channel $m$ only 
for receiving from node $k$ or transmitting to node $j$.
In other words:
\begin{equation}
\sum_{j \in \mathcal{N}, j \neq i} x_{ij}^m + 
\sum_{k \in \mathcal{N}, k \neq i} x_{ki}^m \leq 1 
\,  \, \,  \, \forall \, i \in \mathcal{N} \, , \, \forall \, m \in \mathcal{M}_i
\label{eq:HalfDuplex}
\end{equation}

We use protocol interference model.  
Assume that node $i$ transmits to node $j$
in channel $m$, i.e., $x_{ij}^m = 1$.
Let $p_{ij}^m$ and $g_{ij}^m$ denote the transmission power
and channel gain in channel $m$ of link $ij$. 
Let $P_I$ denote the interference threshold.
$P_I$ should be chosen in such a way so that it is negligible
compared to the noise power $N_0 W$ where $N_0$ is the noise spectral
density. Another node $k$
can transmit to node $h$ in channel $m$ if $p_{kh}^m$
causes negligible interference in node $j$. Hence,
%
%
\begin{equation}
p_{kh}^m \leq \frac{P_I}{g_{kj}^m}  x_{ij}^m
\, \forall (k, h) \in \mathcal{N}, \, k \neq i, \, h \neq j     \label{eq:InterferencePower1}
\end{equation}
Also, a node can allocate power in a link only if it is scheduled, i.e.
\begin{equation}
p_{ij}^m \leq A x_{ij}^m   \,  \, \forall \,  (i \, , j \, \in \mathcal{N}) \, , \, m \in \mathcal{M}
\end{equation}
Here, $A$ is a big number that couples power control and scheduling variables.

\subsubsection{Routing and Link Capacity Constraints}

Let $\mathcal{L}$ be the set of active sessions and $|\mathcal{L}| = L$.
Let $s(l)$, $d(l)$ and $r(l)$ denote the source node, destination node, and
minimum rate requirements of session $l$.  
Let $f_{ij}^m(l)$ denote the flow from node $i$ to node $j$
in channel $m$ for session $l$. If $i$ is the source (destination) node of session $l$, 
the total outgoing (incoming) flow from (to) node $i$
should exceed the minimum rate requirements of session $l$, i.e.,
\begin{equation}
\sum_{j \in \mathcal{N}} \sum_{m \in \mathcal{M}_{ij}} 
f_{ij}^m (l) \geq r(l) \,   \, \, \, \, \, \,
(l \in \mathcal{L} \, , \, i = s(l))
\label{eq:SourceFlow}
\end{equation}
\begin{equation}
\sum_{k \in \mathcal{N}} \sum_{m \in \mathcal{M}_{ki}}  
f_{ki}^m (l) \geq r(l) \,   \, \, \, \, \, \,
(l \in \mathcal{L} \, , \, i = d(l))
\label{eq:DestinationFlow}
\end{equation}
%
The incoming flow
of session $l$ should match the outgoing flow
in an intermediate node $i$:
\begin{equation}
\sum_{j \in \mathcal{N}}^{j \neq s(l)} \sum_{m \in \mathcal{M}_{ij}} f_{ij}^m (l) 
=  \sum_{k \in \mathcal{N}}^{k \neq d(l)} \sum_{m \in \mathcal{M}_{ki}} f_{ki}^m (l) 
\, \, \, \forall \, (l \in \mathcal{L}, \, i \in \mathcal{N}, \, i \neq s(l),d(l)) 
\label{eq:IntermediateFlow}
\end{equation}
Additionally, the aggregated flows of all sessions in a particular
link should not exceed the capacity of the link. Therefore,
\begin{equation}
\sum_{l \in \mathcal{L}}^{s(l) \neq j \, , \, d(l) \neq i} 
f_{ij}^m(l) \leq  c_{ij}^m  \, \, (i \, , j \, \in \mathcal{N} \, , \, i \neq j \, , \, \forall \,  m \, \in \, \mathcal{M}_{ij} \, , \, \mathcal{M}_{ij} \neq \emptyset)
\label{eq:FlowCapacity}
\end{equation}
where
%
%
\begin{eqnarray}
c_{ij}^m & \leq &  W \log (1 + s_{ij}^m)  \,  \, \,  (i \, , j \, \in \mathcal{N} \, , \, i \neq j)  
\label{eq:Capacity}  \\
s_{ij}^m & = & \frac{g_{ij}^m p_{ij}^m}{N_0 W}  , \,  \, \,  (i \, , j \, \in \mathcal{N} \, , \, i \neq j) .
\label{eq:SNR}
\end{eqnarray}

In the above, $c_{ij}^m$ and $s_{ij}^m$ denote the capacity and signal-to-noise-ratio in link $ij$ of channel $m$.
The denominator of $s_{ij}^m$ only contains $N_0 W$ 
because~\eqref{eq:InterferencePower1} ensures
that the interference from other nodes
is negligible compared to the noise power.

The constraints described above resemble previous works that focus on \emph{transmission power} based cross-layer optimization (i.e., power control, scheduling and routing).
The novelty of the work here is in relating the hardware constraints imposed by the radio front-end to the above cross-layer optimization as we describe next.

\subsection{System Power Constraints}

Let $p_{t,i}$ and $p_{r,i}$ denote the system
power consumption in the transmitter and receiver path of node $i$.
The total system power consumption, $P_{tot}$, is:
\begin{equation}
P_{tot} = \sum_{i \in \mathcal{N}} \bigl(p_{t,i} + p_{r,i} \bigr)
\label{eq:TotalPower1}
\end{equation}
Equation~\eqref{eq:SystemPowerTx} and~\eqref{eq:SystemPowerRx}
show a radio node also contains a fixed amount of power
(independent of sampling rate) if it transmits or receives
in a channel.
Denoting $\alpha_{1_i}$ and $\beta_{1_i}$ as this fixed 
power consumption of node $i$'s transmit and receive path
respectively, we find,
\begin{eqnarray}
\alpha_{1_i} & \geq & \alpha_1 x_{ij}^m \, \, \, \, \forall m \in \mathcal{M}_{ij} , \,
 j \in \mathcal{N}, \, i \in \mathcal{N}  \label{eq:AnalogPower1}  \\
\beta_{1_i} & \geq & \beta_1 x_{ki}^m \, \, \, \, \forall m \in \mathcal{M}_{ij} , \,
k \in \mathcal{N} \, i \in \mathcal{N}.
\label{eq:AnalogPower2}
\end{eqnarray}

Using~\eqref{eq:SystemPowerTx},~\eqref{eq:SystemPowerRx},~\eqref{eq:TotalPower1},
~\eqref{eq:AnalogPower1} and~\eqref{eq:AnalogPower2},
\begin{equation}
\sum_{i \in \mathcal{N}} (\alpha_{1_i} + \alpha_2 f_{st,i} + \sum_{j \in \mathcal{N}} \sum_{m \in \mathcal{M}_{ij}} k_{pa} p_{ij}^m
+ \beta_{1_i} + \beta_2 f_{sr,i}) =  P_{tot}   \label{eq:SystemPowerCon}
\end{equation}
where $f_{st,i}$ and $f_{sr,i}$ denote the sampling rates in the transmitter and
receiver of node $i$.

The constraints introduced so far involve channel scheduling decisions and 
sampling rate of different nodes. We now show how channel scheduling decisions influence the sampling rate
of the nodes, which in turn influences the total system power.

\subsection{Relation between Channel Scheduling and System Power}  \label{sec:SpectrumSpan}

The sampling rate of a transceiver depends on its spectrum span, which in turn is determined by the choice of channels 
selected for its intended transmission.  


%
\begin{figure}[t]
\centering
\includegraphics[scale=0.5]{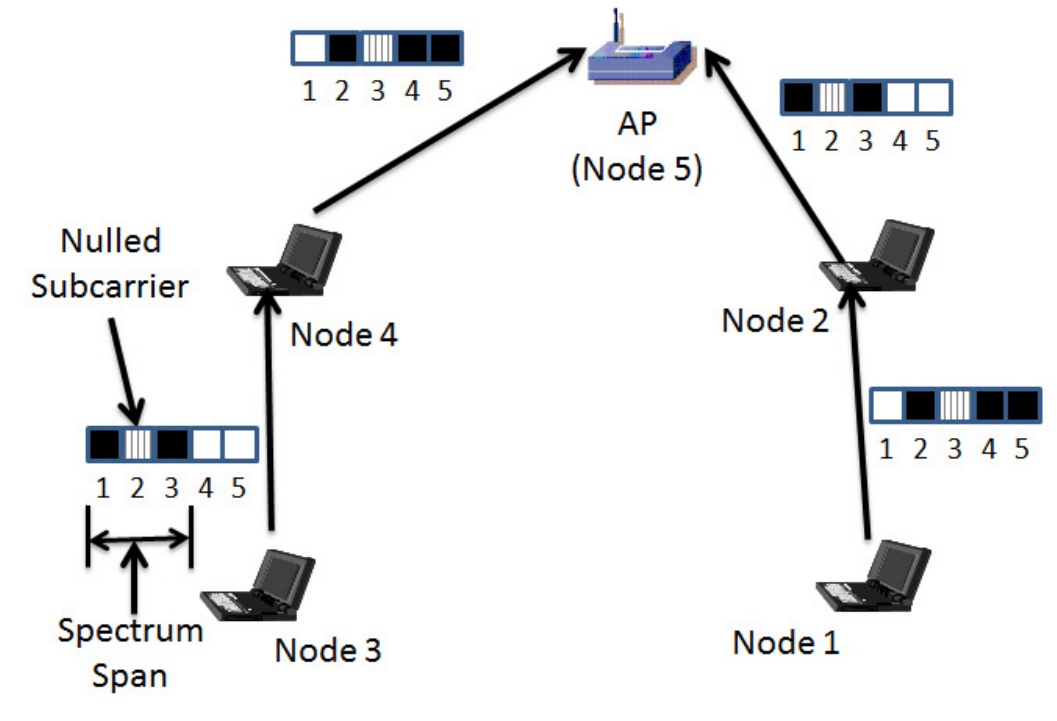}
\caption{Spectrum span, occupied and
nulled subcarrier in an NC-OFDMA based multi-hop network. 
Black solid, lined ash and white boxes denote occupied,
nulled and un-spanned subcarriers respectively.
Spectrum span is the summation of occupied and nulled subcarriers. }
\centering
\label{fig:Spectrum_Span}
\end{figure}

Let $x_i^{t,m}$ and $x_i^{r,m}$ denote the following:
\begin{equation}
  x_{i}^{t,m} =\begin{cases}
    1, & \text{if $i$ transmits to any node $j \, \in \, \mathcal{N}$ in channel $m$}.\\
    0, & \text{otherwise}.   
  \end{cases}   \nonumber
\end{equation} 
\begin{equation}
  x_{i}^{r,m} =\begin{cases}
    1, & \text{if $i$ receives from any node $j \, \in \, \mathcal{N}$ in channel $m$}.\\
    0, & \text{otherwise}.
  \end{cases}    \nonumber 
\end{equation} 
In other words,
\begin{eqnarray}
x_i^{t,m} & \geq & x_{ij}^m  \, \, \forall \, j \, \in \, \mathcal{N}, \nonumber \\
x_i^{r,m} & \geq & x_{ki}^m  \, \, \forall \, k \, \in \, \mathcal{N}, \label{eq:SchedulingAllNode}
\end{eqnarray}
Using this notation, analog power equations of~\eqref{eq:AnalogPower1} and~\eqref{eq:AnalogPower2}
are redefined as,
\begin{equation}
\alpha_{1_i} \geq \alpha_1  x_i^{t,m} \, , \, \, \, \, 
\beta_{1_i} \geq \beta_1  x_i^{r,m} , \, \, \forall m \in \mathcal{M}_i , \,
\, \forall i \in \mathcal{N}
\label{eq:FixScheduler}
\end{equation}
\emph{We define spectrum span as the gap between the furthest edges 
of the used channels}.
Node $i$'s uppermost used channel index in the transmitter path is:
\begin{equation}
\max_{m \in \mathcal{M}_i} m \cdot x_i^{t,m}
\label{eq:UpperChannelIndex}
\end{equation}
Node $i$'s lowermost used channel index in the transmitter path is:
\begin{equation}
\min_{m \in \mathcal{M}_i} (m \cdot x_i^{t,m} + |M| \cdot (1 - x_i^{t,m})) .
\label{eq:LowerChannelIndex}
\end{equation}
%


The second term of~\eqref{eq:LowerChannelIndex} ensures that 
we do not consider the index $i$'s for which $x_i^{t,m} = 0$.
Let $q_{t,i}$ and $q_{r,i}$ denote the spectrum spans of the transmitter
and receiver path of node $i$. Now,
\begin{equation}
q_{t,i} = W \cdot \max \bigl(\bigl(\max_{m \in \mathcal{M}_i} \bigl(m \cdot x_i^{t,m} \bigr)  
- \min_{m \in \mathcal{M}_i} \bigl(m \cdot x_i^{t,m} + |M| \cdot (1 - x_i^{t,m}) \bigr) + 1 \bigr), 0 \bigr) 
\label{eq:SpanTx}
\end{equation}
%
\begin{equation}
q_{r,i} =   W \cdot \max \bigl(\bigl(\max_{m \in \mathcal{M}_i} \bigl(m \cdot x_i^{r,m} \bigr) \\ 
- \min_{m \in \mathcal{M}_i} \bigl(m \cdot x_i^{r,m} + |M| \cdot (1 - x_i^{r,m}) \bigr) + 1 \bigr) , 0 \bigr)
\label{eq:SpanRx}
\end{equation}

Fig.~\ref{fig:Spectrum_Span} illustrates the concept of spectrum span
in a five node multi-hop network. We now verify~\eqref{eq:UpperChannelIndex} -~\eqref{eq:SpanRx}
by focusing on the spectrum span of node $3$ of Fig.~\ref{fig:Spectrum_Span}. 
There are $5$ available channels. Node $3$ transmits in channel $1$ and $3$, and nulls channel $2$.
Node $3$ does not receive in any channel from other nodes.
Hence,
\begin{equation}
\mathcal{M} = \{1, 2, 3, 4, 5 \} \, , \, x_3^{r,m} = 0 \, \forall \, m \, \in \, \mathcal{M},  \nonumber
\end{equation}
\begin{equation}
x_3^{t,m} = 1 \, \forall m \, \in \, \{1, 3 \} \, , \, x_3^{t,m} = 0 \, \forall m \, \in \, \{2, 4, 5 \}  \nonumber
\end{equation}
Using the values of $\mathcal{M}$, $x_3^{r,m}$ and $x_3^{t,m}$ in~\eqref{eq:SpanTx} and~\eqref{eq:SpanRx},
$q_{r,3} = 0$ and $q_{t,3} = 3W$. Fig.~\ref{fig:Spectrum_Span} shows that
node $3$ does not receive from any node and spans three channels while transmitting to node $4$.

Now, using~\eqref{eq:FixScheduler},~\eqref{eq:SpanTx} and~\eqref{eq:SpanRx}, along with the fact that sampling rate
of a transceiver should be at least twice its spectrum span, 
we convert~\eqref{eq:SystemPowerCon} to the following:
\begin{equation}
\sum_{i \in \mathcal{N}} (\alpha_{1_i} + 2 \alpha_2 q_{t,i} + \sum_{j \in \mathcal{N}} \sum_{m \in \mathcal{M}_{ij}} p_{ij}^m
+ \beta_{1_i} + 2 \beta_2 q_{r,i}) =  P_{tot}
\label{eq:TotPowerWithSpan}
\end{equation}
Equation~\eqref{eq:FixScheduler},~\eqref{eq:SpanTx}-~\eqref{eq:TotPowerWithSpan} relate 
total system power to the scheduling variables.

In this work, our objective is to minimize the total system power subject to rate
requirements. Using the above equations, we formulate our optimization
problem and show it in~\eqref{fig:ModifiedProblem}.
%

\begin{subequations}
\label{fig:ModifiedProblem}
\begin{equation}
\min \, \, \,   P_{tot}
\label{eq:ModifiedObjective}
\end{equation}
\begin{equation}
\sum_{j \in \mathcal{N}} \sum_{m \in \mathcal{M}_{ij}} 
f_{ij}^m (l) \geq r(l) \,   \, \, \, \, \, \,
(l \in \mathcal{L} \, , \, i = s(l))  \, \, , \, \,
\sum_{k \in \mathcal{N}} \sum_{m \in \mathcal{M}_{ki}}  
f_{ki}^m (l) \geq r(l) \,   \, \, \, \, \, \,
(l \in \mathcal{L} \, , \, i = d(l))
\end{equation}
\begin{equation}
\sum_{j \in \mathcal{N}}^{j \neq s(l)} \sum_{m \in \mathcal{M}_{ij}} f_{ij}^m (l) 
 =  \sum_{k \in \mathcal{N}}^{k \neq d(l)} \sum_{m \in \mathcal{M}_{ki}} f_{ki}^m (l) \, \,
 (l \in \mathcal{L}, \, i \in \mathcal{N}, \, i \neq s(l),d(l)) 
\end{equation}
\begin{equation}
\sum_{l \in \mathcal{L}}^{s(l) \neq j \, , \, d(l) \neq i} 
f_{ij}^m(l) \leq  c_{ij}^m  \, \, (i \, , j \, \in \mathcal{N} \, , \, i \neq j \, , \, \mathcal{M}_{ij} \neq \emptyset)
\end{equation}
\begin{equation}
c_{ij}^m \leq  W \log (1 + s_{ij}^m)  \,  \, \,  (i \, , j \, \in \mathcal{N} \, , \, i \neq j)    \, \, , \, \,
s_{ij}^m  =  \frac{g_{ij}^m p_{ij}^m}{N_0 W}  , \,  \, \,  (i \, , j \, \in \mathcal{N} \, , \, i \neq j)
\label{eq:CapacitySINR}
\end{equation}
\begin{equation}
p_{kh}^m \leq  \frac{P_I}{g_{kj}} x_{ij}^m
\, \forall \, k \in \mathcal{N}, \, h \in \mathcal{N}, \, k \neq h   \, \, ,
\, \, p_{ij}^m \leq A x_{ij}^m   \,  \, \forall \,  (i \, , j \, \in \mathcal{N}) \, , \, m \in \mathcal{M}
\end{equation}
\begin{equation}
q_{t,i} \geq W \cdot \bigl(\max_{m \in \mathcal{M}_i} \bigl(m \cdot x_i^{t,m} \bigr) \\ 
- \min_{m \in \mathcal{M}_i} \bigl(m \cdot x_i^{t,m} + |M| \cdot (1 - x_i^{t,m}) \bigr) + 1 \bigr) 
\label{eq:ModifiedSpanTx}
\end{equation}
\begin{equation}
q_{r,i} \geq   W \cdot \bigl(\max_{m \in \mathcal{M}_i} \bigl(m \cdot x_i^{r,m} \bigr) \\ 
- \min_{m \in \mathcal{M}_i} \bigl(m \cdot x_i^{r,m} + |M| \cdot (1 - x_i^{r,m}) \bigr) + 1 \bigr) 
\label{eq:ModifiedSpanRx}
\end{equation}
\begin{equation}
\alpha_{1_i} \geq \alpha_1  x_i^{t,m} \, , \, \, \, \, 
\beta_{1_i} \geq \beta_1  x_i^{r,m}  \, \, \forall m \in \mathcal{M}_i 
\, \forall i \in \mathcal{N} 
\end{equation}
\begin{equation}
\sum_{j \in \mathcal{N} \, , j \neq i} x_{ij}^m + 
\sum_{k \in \mathcal{N} \, , k \neq i} x_{ki}^m \leq 1 
\,  \, \,  \, \forall \, i \in \mathcal{N} \, , \, \forall \, m \in \mathcal{M}_i
\end{equation}
%
%
%
\begin{equation}
\sum_{i \in \mathcal{N}} \bigl(\alpha_{1_i} + 2 \alpha_2 q_{t,i}  + \sum_{j \in \mathcal{N}} \sum_{m \in \mathcal{M}_{ij}} p_{ij}^m
+ \beta_{1_i} + 2 \beta_2 q_{r,i} \bigr) \leq  P_{tot}
\label{eq:ModifiedSystemPower}
\end{equation}
\begin{equation}
x_{ij}^m \in \{0,1\} \, , \, s_{ij}^m \geq 0 \, \, (i, j \in \mathcal{N}, \, i \neq j, \, 
m \in \mathcal{M}_{ij} ) \, , \, \, q_{t,i} \geq 0, \, q_{r,i} \geq 0 \, \, 
\forall i \in \mathcal{N}
\label{eq:ModifiedVar1}
\end{equation}
\begin{equation}
P_{tot} \, , f_{ij}^m (l) \geq 0 \, \bigl(l \in \mathcal{L}, m \in \mathcal{M}_{ij},
 \, i,j \in \mathcal{N}, \, i \neq j, \, i \neq d(l), \, j \neq s(l), \, \mathcal{M}_{ij} \neq \emptyset  \bigr)
\label{eq:ModifiedVar2}
\end{equation}
\end{subequations}

Next two sections provide solution methodologies
of~\eqref{fig:ModifiedProblem}.

\section{Branch-and-Bound based Solution Overview}  \label{sec:Solution}

\subsection{Linearization of the Optimization Problem}

The optimization problem
of~\eqref{fig:ModifiedProblem} is a mixed integer non-linear 
program (MINLP). The non-linearity of~\eqref{fig:ModifiedProblem}
comes from the logarithm function of~\eqref{eq:CapacitySINR}
and max-min functions of~\eqref{eq:ModifiedSpanTx}
and~\eqref{eq:ModifiedSpanRx}.

The logarithmic function of the capacity term in~\eqref{eq:CapacitySINR}
can be linearized by reformulation linearization technique (RLT)~\cite{RLT}.
RLT uses a number of tangential supports at different
points of the logarithmic curve and generates a convex hull linear relaxation
of the logarithmic function. We ask interested readers to go through~\cite{RLT} 
for the details of RLT.

The max-min equations 
of~\eqref{eq:SpanTx} and~\eqref{eq:SpanRx} can be re-written in following linear forms:
\begin{equation}
q_{t,i}  + W \bigl(m_2 \cdot x_i^{t,m_2} + |M| \cdot (1 - x_i^{t,m_2}) \bigr) 
\geq W  \bigl( m_1 \cdot x_i^{t,m_1} + 1 \bigr)  \, \, \forall (m_1,m_2) \in \mathcal{M}_i 
\, , \, \forall i \in \mathcal{N}
\end{equation}
\begin{equation}
q_{r,i}  + W \bigl(m_2 \cdot x_i^{r,m_2} + |M| \cdot (1 - x_i^{r,m_2}) \bigr) \\
\geq W  \bigl( m_1 \cdot x_i^{r,m_1} + 1 \bigr) \, \,  \, \, \forall (m_1,m_2) \in \mathcal{M}_i 
\, , \, \forall i \in \mathcal{N}
\end{equation}
\begin{equation}
q_{t,i} \geq 0 \, , \, \, q_{r,i} \geq 0 \, \, \, \forall i \in \mathcal{N}
\end{equation}
The optimization problem of~\eqref{fig:ModifiedProblem} with the reformulated
linear equations can be directly solved in CVX~\cite{CVX} (with MOSEK~\cite{Mosek}) software.
This problem is a MILP. 
CVX uses branch-and-bound algorithm~\cite{RLT} to solve this problem.

\subsection{Feasible Solution}

CVX outputs flow variables $f_{ij}^m(l)$,
scheduling decisions $x_{ij}^m$ and power variables $p_{ij}^m$
$\bigl(l \in \mathcal{L}, m \in \mathcal{M}_{ij},
 \, i,j \in \mathcal{N}, \, i \neq j,   \bigr)$.
Since we relaxed flow capacity equations to get this output,
the resultant flow rates may exceed the capacity of the links.
We keep flow variables and scheduling decisions
unperturbed and increase power variables to get feasible solutions.
We use the following set of equations for flows and powers:
\begin{eqnarray}
\sum_{i \in \mathcal{L}} f_{ij}^m (l) & = &
W log(1 + \frac{p_{ij}^m g_{ij}^m}{N_0 W}) \nonumber  \\
p_{ij}^m & = & \frac{N_0 W}{g_{ij}^m} 
\bigl[exp\{\frac{\sum_{i \in \mathcal{L}} f_{ij}^m (l)}{W}\}-1\bigr] \, \, \, \forall m \in \mathcal{M}_{ij},
 \, i,j \in \mathcal{N}, \, i \neq j.
\end{eqnarray}
These power variables along with flow and scheduling decisions 
of the CVX output form a feasible solution.
We refer to this solution as ``BnBSysPowerMin".
The resulting algorithm however suffers from
exponential complexity in the worst case scenario.
Next section develops low complexity algorithms
to minimize system power in a multi-hop network, i.e., 
to solve the original optimization problem of~\eqref{fig:ModifiedProblem}.

\section{Low Complexity Algorithm Design}  \label{sec:Insights}

We first focus on the system power minimization of a point-to-point
link. Thereafter, we use the insights obtained from this scenario
to develop a low complexity algorithm to minimize system power in a multi-hop network.

\subsection{Theoretical Insights from Point-to-Point Link Case}

In a point-to-point link, the optimization problem of~\eqref{fig:ModifiedProblem}
takes the following form:

\begin{subequations}
\begin{equation}
\min \, \, \, \, \sum_{m \in \mathcal{M}} p^m 
+ \alpha_1 + 2 \alpha_2 q + \beta + 2 \beta_2 q
\label{eq:SinglePairProblemJoint}
\end{equation}
\begin{equation}
s.t. \, \, \, \, q \geq W \bigl(\max_{m \in \mathcal{M}} (m \cdot x^m)
- \min_{m \in \mathcal{M}} (m \cdot x^m + |M| (1 - x^m)) + 1 \bigr)
\, \, , \, \, x^m \in \{0,1\} \, , \, \forall m \in \mathcal{M} \, , \, q \geq 0
\label{eq:ScheduleConstraintJoint}
\end{equation}
\begin{equation}
\sum_{m \in \mathcal{M}} W \log_2 \bigl(1 + \frac{p^m g^m}{N_0 W} \bigr) \geq r
\, \, , \, \, p^m \geq 0, \, \, \forall m \in \mathcal{M}
\label{eq:PowerConstraintJoint}
\end{equation}
\begin{equation}
p^m \leq A x^m  \, \, \forall m \in \mathcal{M}
\label{eq:CouplingConstraintJoint}
\end{equation}
\label{eq:P2PMinTxCkt}
\end{subequations}

In above equations, $p^m$ and $g^m$ denote the allotted power and link
gain in channel $m$, respectively. 
Rate requirement and spectrum span are denoted by $r$ and $q$, respectively.
In the objective function, $\sum_{m \mathcal{M}} p^m$ denotes transmit power and $\alpha_1 + 2 \alpha_2 q + \beta + 2 \beta_2 q$ denotes circuit power
consumption.

The optimization problem of~\eqref{eq:P2PMinTxCkt} is the combination of two separate 
optimization problems. The objective is to minimize
the summation of transmit and circuit power.
Eq.~\ref{eq:ScheduleConstraintJoint} denotes the constraints associated
with circuit power minimization problem and it only involves spectrum span and scheduling
variables. Eq.~\ref{eq:PowerConstraintJoint} denotes the constraints
associated with transmit power minimization problem and it only involves power variables.
Eq.~\ref{eq:CouplingConstraintJoint} relates the power and scheduling variables and couples
these two optimization problems.

The optimization problem of~\eqref{eq:P2PMinTxCkt} has two sub cases.
These sub-cases depend on the values of $\alpha_2$ and $\beta_2$, i.e., the
slope of ADC \& DAC's power consumption versus sampling rate curves 

\subsubsection{Case I: Transmit Power Minimization}

When $\alpha_2$ and $\beta_2$ are very small, i.e.,  
ADC and DAC's power consumption versus sampling rate curves are very flat, 
the impact of spectrum
span on system power becomes negligible. 
Scheduling decisions do not influence system power
that much and we can
concentrate on minimizing transmit power. 
Then the optimization problem gets reduced to:

\begin{subequations}
\begin{equation}
\min \sum_{m \in \mathcal{M}} p^m
\label{eq:TxPowerObjective}
\end{equation}
\begin{equation}
\sum_{m \in \mathcal{M}} W \log_2 \bigl(1 + \frac{p^m g^m}{N_0 W} \bigr) \geq r
\, \, , p^m \leq Ax^m  \, \, , p^m \geq 0, \, \, x^m \in \{0,1\} \, \, \forall m \in \mathcal{M}
\label{eq:PowerConstraint}
\end{equation}
\label{eq:P2PMinTx}
\end{subequations}

Since $x^m$'s are not present in the objective function,
we can just solve the problem with $p^m$ variables and then enforce $x^m = 1$
for every positive $p^m$.

The optimization problem of~\eqref{eq:P2PMinTx} is similar to the classical waterfilling~\cite{Cover}
problem, which maximizes rate subject to a total power constraint. 
In the remainder of the paper, we refer to the above problem `TxPowerMin'.
The solution to this problem selects the ``good" channels in the network and spreads
power across the whole spectrum~\cite{Cover}. 

\subsubsection{Case II: Circuit Power Minimization}

When $\alpha_2$ and $\beta_2$ are very large, i.e., ADC and DAC's power consumption
are very steep, circuit power consumption dominates system power. 
Transmit power variables do not
influence circuit power that much and we can just concentrate on
minimizing circuit power. The optimization problem reduces to,

\begin{subequations}
\begin{equation}
\min \, \, \, \, \alpha_1 + 2 \alpha_2 q + \beta + 2 \beta_2 q
\label{eq:SinglePairProblem}
\end{equation}
\begin{equation}
s.t. \, \, \, \, q \geq W \bigl(\max_{m \in \mathcal{M}} (m \cdot x^m)
- \min_{m \in \mathcal{M}} (m \cdot x^m + |M| (1 - x^m)) + 1 \bigr)
\, \, , \, \, x^m \in \{0,1\} \, , \, \forall m \in \mathcal{M} \, , \, q \geq 0
\label{eq:ScheduleConstraint}
\end{equation}
\begin{equation}
\sum_{m \in \mathcal{M}} W \log_2 \bigl(1 + \frac{p^m g^m}{N_0 W} \bigr) \geq r
\, \, , \, \, p^m \geq 0, \, \, \forall m \in \mathcal{M}
\, \, \, , \, \, \, p^m \leq A x^m  \, \, \forall m \in \mathcal{M}
\label{eq:PowerConstraint}
\end{equation}
%
%
\label{eq:P2PMinCkt}
\end{subequations}

The objective of the optimization problem of~\eqref{eq:P2PMinCkt} increases with spectrum span $q$.
Minimum circuit power occurs if we schedule only one channel
 and allocate enough power in that channel so that it can 
 meet rate requirement. Scheduling any single channel leads to 
same circuit power in the above optimization problem.
Since the original system power minimization problem contains both
transmit and circuit powers, it is prudent to select the channel with the best link gain.
This greedy selection requires less transmit power to meet rate requirement.

\subsubsection{Trade-off between transmit and circuit power minimization}

If ADC and DAC's power consumption versus sampling rate curves 
are very flat, our algorithm
selects all good channels in the network.
If ADC and DAC's power consumption vs. sampling rate 
curves are very steep, our algorithm
selects the channel with the best link gain.
In a practical setting with commercial ADCs and DAC, 
our algorithm trades off between transmit and circuit power.

\subsection{Polynomial Time Algorithm for a Multihop Networks}  \label{sec:GreedyAlgo}

The analysis from the point-to-point link case provides us insights
to develop a low complexity greedy algorithm to solve the optimization problem of~\eqref{fig:ModifiedProblem}. 
The algorithm minimizes the circuit power by using only one channel at the first step
and then tries to minimize the system power, i.e., the combination
of transmit and circuit power, by greedily selecting more and
more channels in the subsequent steps.
Our greedy algorithm can be explained simply as follows:

\begin{itemize}

\item Find the initial route between each sender and receiver using the shortest 
path algorithm~\cite{Algorithm}.

\item Assign the best channel to each link unless the current assignment
interferes with previously assigned channels.

\item For each active link, check if any other channel assignment
reduces system power.

\item Once channels are scheduled, determine power allocation and routing 
through a convex optimization program.

\end{itemize}

Our greedy system power minimization
algorithm consists of the central program of Algorithm~\ref{alg:GreedyMultiHop}
and the sub-routines of Algorithm~\ref{alg:GreedyScheduling} and 
Algorithm~\ref{alg:InterferenceCheck}. Next three sub-sections 
describe the central program and the sub-routines.
We analyze the complexity of our algorithm in 
Sec.~\ref{sec:Complexity}.

\subsubsection{Central Program}

\begin{algorithm}
\caption{Polynomial Time Algorithm to Minimize System Power in a 
Multi-hop Network}
\label{alg:GreedyMultiHop}
\SetKwInOut{Input}{Input}
\SetKwInOut{Output}{Output}
\Input{$\mathcal{M}$, $s(l), \, d(l), \, r(l) \, \forall l \in \mathcal{L}$, $W$, $N_0$, $g_m \, \forall m \in \mathcal{M}$}
\Output{$x_m , \, p_m \forall m \in \mathcal{M}$ , $P_{tot}$ , $f_{ij}^m(l) \, \forall m \in \mathcal{M}, \, \forall l \in \mathcal{L}, \, \forall (i,j) \, \in \mathcal{E}$}
\begin{algorithmic}[1]
\STATE Denote $g_{ij}$ as the average gain (e.g. path loss plus shadowing)
of link $ij$.    \label{GM:Op1}   
\STATE Assign weight $w_{ij}$ to each link, $w_{ij} = \frac{1}{g_{ij}}$. \label{GM:Op2}
\STATE Find shortest path between between source $(s(l))$ and destination $(d(l))$
of every session $l \in \mathcal{L}$ based on the link weight $w$. \label{GM:Op3}
\STATE $b_{ij}(l) = 1$ if link $ij$ falls in the routing path of any 
session $l \in \mathcal{L} $. \label{GM:Op4}
\STATE A link is active if it falls in the routing path of any session, i.e.,
$x_{ij} = 1$ if $ \exists l \in \mathcal{L}$ s.t. $b_{ij}(l) = 1$. \label{GM:Op5}
\STATE Define $\mathcal{A}$ to be set of active links. \label{GM:Op6}
\STATE Flow in each link, $f_{ij} = \sum_{l \in \mathcal{L}} b_{ij}(l) r(l)$.  \label{GM:Op7}
\STATE $P_{tot} = \infty$, $x_{ij}^m = p_{ij}^m = 0 \, \forall \, 
(i, j) \, \in \, \mathcal{E} \, , \, \forall m \, \in \, \mathcal{M} $ ,
$x_i^{t,m} = x_i^{r,m} = 0, \, 
\alpha_{1_i} = \beta_{1_i} = 0, \forall \, i \, \in \mathcal{N}, 
\forall m \in \mathcal{M}$  \label{GM:Op8}
\WHILE{True}   \label{GM:Op9}
\STATE $flag = 0$ ; \label{GM:Op10}
\FOR{$(a,b) \in \mathcal{A}$}   \label{GM:Op11}
\STATE $\bigl(flag , x_{ij}^m , p_{ij}^m \,
 \forall \,  m \, \in \, \mathcal{M}  \, , \,
(i, j) \, \in \, \mathcal{E}  \bigr) = $   \\
GreedyAlgo$\bigl(\mathcal{M}, (a,b), f_{ab}, r, W, N_0, flag, P_{tot}, \alpha_{1_i}, \beta_{1_i} ,
x_i^{t,m} , x_i^{r,m}, x_{ij}^m , p_{ij}^m, g_{ij}^m \, \forall \,  m \, \in \, \mathcal{M} \, ,
 \, i \, \in \, \mathcal{N} \, , (i, j) \, \in \, \mathcal{E}   \bigr)$    \label{GM:Op13}
\ENDFOR   \label{GM:Op14}
\IF{$flag = |A|$}   \label{GM:Op15}
\STATE Break.  \label{GM:Op16}
\ENDIF  \label{GM:Op17}
\ENDWHILE  \label{GM:Op18}
\STATE Solve the optimization problem of~\eqref{fig:ModifiedProblem} where
scheduling variables ($x_{ij}^m \, \forall \, m \, \in \, \mathcal{M} \, , \,
\forall \, (i, j) \, \in \, \mathcal{E}$)
are constants, not variables. Find power
allocation, scheduling and routing variables and total power, $P_{tot}$.  \label{GM:Op19}
\end{algorithmic}
\end{algorithm}

Algorithm~\ref{alg:GreedyMultiHop} shows the pseudocode of our
greedy polynomial time algorithm. Operation~\ref{GM:Op1} finds large scale gains
of all links by averaging small scale fading in time
or frequency domain.
Operation~\ref{GM:Op2} assigns weight to each link. 
Operation~\ref{GM:Op3} finds the shortest path between each sender and forwarder
based on the assigned weights of operation~\ref{GM:Op2}.
Operation~\ref{GM:Op5} finds the active links and 
operation~\ref{GM:Op7} calculates the flow
requirement among these links. 
Operation~\ref{GM:Op8} initiates total power ($P_{tot}$), power allocation 
($p_{ij}^m)$ and scheduling ($x_{ij}^m$) variables to zero
for the greedy channel scheduling algorithm. We initiate an outer loop
in operation~\ref{GM:Op9}.
Operation~\ref{GM:Op11}-\ref{GM:Op14} calls the subroutine of Algorithm~\ref{alg:GreedyScheduling}
and checks if any link should be assigned a channel.
The outer loop breaks if none of the 
active links becomes suitable to be assigned a channel. 
This outer loop 
determines the channel scheduling ($x_{ij}^m$) variables. 

We obtain power allocation ($p_{ij}^m$)
and routing path ($f_{ij}(l)$) variables from the 
optimization problem of~\eqref{fig:ModifiedProblem}
where scheduling variables ($x_{ij}^m$) are constants.
Since we fix the integer variables of~\eqref{fig:ModifiedProblem},
the total power minimization problem becomes a convex optimization
program and can be solved in polynomial time~\cite{Boyd}.

We assume one path (shortest path), i.e., no flow splitting, per session
during the initial routing topology design of operation~\ref{GM:Op1}-~\ref{GM:Op3}.
This allows us to easily calculate the flow requirement of
each link which we later use in the greedy scheduling algorithm.
We consider optimal flow splitting, based on the selected
scheduling variables, in the final 
optimization of operation~\ref{GM:Op19} of Algorithm~\ref{alg:GreedyMultiHop}.

\subsubsection{Greedy Scheduling Algorithm}

The greedy scheduling algorithm of Algorithm~\ref{alg:GreedyScheduling}
is a sub-routine that's called from operation~\ref{GM:Op13} of the central program of 
Algorithm~\ref{alg:GreedyMultiHop}. The sub-routine receives 
previously assigned scheduling and power allocation
variables from the central program.
The central controller also asks the sub-routine to focus on a 
particular link $(a,b)$.
The sub-routine iterates through all
available channels and finds the best available
channel for $(a,b)$.

Operation~\ref{GS:Op1} stores the global scheduling ($x_{ij}^m$) 
and power allocation ($p_{ij}^m$) variables in local 
dummy variables $x_{new,ij}^m$ and $p_{new,ij}^m$ respectively.
Operation~\ref{GS:Op2} of Algorithm~\ref{alg:GreedyScheduling} starts the iteration
for all channels. 
Operation~\ref{GS:Op3} assigns the current channel to the focus link $(a,b)$.
Operation~\ref{GS:Op4} and~\ref{GS:Op5} calculate the transmit span ($q_{t,i}$) and 
receiver span ($q_{r,i}$) for all nodes $i \in \mathcal{N}$.
Operation~\ref{GS:Op6} assumes equal flow allocation among the selected 
channels and finds the power allocation in link $(a,b)$ to meet rate requirement.
Operation~\ref{GS:Op7} calls the sub-routine of Algorithm~\ref{alg:InterferenceCheck}
and checks if current channel assignment causes interference to
other links. 
Operation~\ref{GS:Op11} calculates total system power of the current iteration.
Operation~\ref{GS:Op14} comes out of the loop.
Operation~\ref{GS:Op15} finds the minimum system power
($P_{tot,new}$) among all possible channel allocations. 
Operation~\ref{GS:Op16}-\ref{GS:Op22} compare 
the achievable minimum system power ($P_{tot,new}$) with the stored 
global system power ($P_{tot}$)
and update scheduling and power variables accordingly.

\begin{algorithm}
\caption{Greedy Scheduling Algorithm}
\label{alg:GreedyScheduling}
\SetKwInOut{Input}{Input}
\SetKwInOut{Output}{Output}
\Input{$\mathcal{M}$, Link $(a,b)$, $f_{ab}$,  $r$, $W$, $N_0$, $flag$,
$\alpha_{1_i}$, $\beta_{1_i}$, $x_i^{t,m}$, $x_i^{r,m}$, 
$x_{ij}^m \, , \, p_{ij}^m \, , \, g_{ij}^m \, \forall \, 
(i, j) \, \in \, \mathcal{E} \, , \, \forall \, i \, \in \, \mathcal{N} \, , \, \forall m \, \in \, \mathcal{M} $}
\Output{$flag$, $x_{ij}^m \, , \, \forall \, 
(i, j) \, \in \, \mathcal{E} \, , \, \forall m \, \in \, \mathcal{M} $}
\begin{algorithmic}[1]
\STATE $x_{new,ij}^m = x_{ij}^m \, , \, p_{new,ij}^m = p_{ij}^m \, ,
\forall \, (i, j) \, \in \, \mathcal{E} \, , \, \forall m \, \in \, \mathcal{M} $ \label{GS:Op1}

\STATE $f_{new,ab}^m = \frac{f_{ab}}{\sum_{m \in \mathcal{M}} x_{new,ab}^m + 1} \cdot x_{new,ab}^m$ , $p_{new,ab}^m = \frac{N_0 W}{g_{ab}^m} \cdot \bigl(2^{\frac{f_{new,ab}^m}{W}} - 1  \bigr) 
x_{new,ab}^m \, \forall \, m \, \in \, \mathcal{M} $ \label{GS:Op155}
\FOR{$m \in \mathcal{M}$ where $x_{ab}^m \neq 1 $}  \label{GS:Op2}
\STATE  $x_{new,ab}^m = 1$. $x_a^{t,m} = 1$. $x_b^{r,m} = 1$. $\alpha_{1_a} = \alpha_1$. $\beta_{1_b} = \beta_1$ \label{GS:Op3}
\STATE $ q_{t,i} = W \cdot \bigl(\max_{m \in \mathcal{M}} \bigl(m \cdot x_i^{t,m} \bigr)  - \min_{m \in \mathcal{M}} \bigl(m \cdot x_i^{t,m} + |M| \cdot (1 - x_i^{t,m}) \bigr) + 1 \bigr) \, \forall \, i \, \in \, \mathcal{N} $   \label{GS:Op4}
\STATE $q_{r,i} =   W \cdot \bigl(\max_{m \in \mathcal{M}_i} \bigl(m \cdot x_i^{r,m} \bigr)  
- \min_{m \in \mathcal{M}_i} \bigl(m \cdot x_i^{r,m} + |M| \cdot (1 - x_i^{r,m}) \bigr) + 1 \bigr)  \, \forall \, i \, \in \, \mathcal{N}  $  \label{GS:Op5}
\STATE $f_{new,ab}^m = \frac{f_{ab}}{\sum_{m \in \mathcal{M}} x_{new,ab}^m} \cdot x_{new,ab}^m$ , $p_{new,ab}^m = \frac{N_0 W}{g_{ab}^m} \cdot \bigl(2^{\frac{f_{new,ab}^m}{W}} - 1  \bigr) 
x_{new,ab}^m  $ \label{GS:Op6}
\STATE $IntFlag =$ IntCheck$\bigl(x_{new,ab}^m, p_{new,ab}^m,
\mathcal{M}, \mathcal{A}, W, N_0, x_{ij}^m, p_{ij}^m \, 
\forall \, (i,j) \, \in \, \mathcal{E}    \bigr)$    \label{GS:Op7}
\IF{IntCheck($\cdot$) = 1}  \label{GS:Op8}
\STATE $P_{new,tot}^m = \infty$  \label{GS:Op9}
\ELSE  \label{GS:Op10}
\STATE $ P_{new,tot}^m  = 
\sum_{i \in \mathcal{N}} \bigl(\alpha_{1_i} + 2 \alpha_2 q_{t,i}  + \sum_{j \in \mathcal{N}} \sum_{m \in \mathcal{M}_{ij}} p_{new,ij}^m
+ \beta_{1_i} + 2 \beta_2 q_{r,i} \bigr)$  \label{GS:Op11}
\ENDIF  \label{GS:Op12}
\STATE $x_a^{t,m} = 0$, $x_b^{r,m} = 0$   \label{GS:Op13}
\ENDFOR \label{GS:Op14}
\STATE $P_{tot,new} = \min_{m \in \mathcal{M}} P_{tot,new}^m$. 
$ind = \arg \min_{m \in \mathcal{M}} P_{tot,new}^m$   \label{GS:Op15}
\IF{$P_{tot,new} < P_{tot}$} \label{GS:Op16}
\STATE $x_{ab}^{ind} = 1$.  \label{GS:Op17}
\STATE $x_a^{t,ind} = 1$. $x_b^{r,ind} = 1$. $\alpha_{1_a} = \alpha_1$. $\beta_{1_b} = \beta_1$.  \label{GS:Op18}
\STATE $f_{ab}^m = \frac{f_{ab}}{\sum_{m \in \mathcal{M}} x_{ab}^m} \cdot x_{ab}^m$ , $p_{ab}^m = \frac{N_0 W}{g_{ab}^m} \cdot \bigl(2^{\frac{f}{W}} - 1  \bigr) x_{ab}^m
\, \forall \, m \, \in \, \mathcal{M} $   \label{GS:Op19}
\ELSE  \label{GS:Op20}
\STATE $flag = flag + 1$  \label{GS:Op21}
\ENDIF  \label{GS:Op22}
\RETURN \label{GS:Op23}
\end{algorithmic}
\end{algorithm}

We assume equal flow per channel in each link
in operation~\ref{GS:Op6} and~\ref{GS:Op19} of Algorithm~\ref{alg:GreedyScheduling}.
This simplifies the calculation of transmission power per channel and 
avoids the computation of a convex optimization program in each loop.
However, we consider optimal flow and power allocation
in our final optimization problem of operation~\ref{GM:Op19} of Algorithm~\ref{alg:GreedyMultiHop}.


\subsubsection{Interference Checking Algorithm}

The sub-routine of Algorithm~\ref{alg:InterferenceCheck} gets called
from operation~\ref{GS:Op7} of the greedy scheduling algorithm of Algorithm~\ref{alg:GreedyScheduling}.
This sub-routine checks if the scheduling and power allocation
of link $ab$ in channel $m$, calculated in Algorithm~\ref{alg:GreedyScheduling}, interferes with
other links. Operation~\ref{Op:IC3} of Algorithm~\ref{alg:InterferenceCheck} 
checks if transmitted power in link $ab$ causes
interference to any non-adjacent link that uses channel $m$.
Operation~\ref{Op:IC6} checks if link $ab$ maintains half duplex relationships
with its adjacent links. Operation~\ref{Op:IC11} updates the interference flag and returns this value
to the greedy scheduling sub-routine of Algorithm~\ref{alg:GreedyScheduling}.

\begin{algorithm}
\caption{Primary and Secondary Interference Checking Algorithm}
\label{alg:InterferenceCheck}
\SetKwInOut{Input}{Input}
\SetKwInOut{Output}{Output}
\Input{$x_{new,ab}^m$, $p_{new,ab}^m$, $\mathcal{M}$, $\mathcal{A}$, 
$W$, $N_0$, $x_{ij}^m$, $p_{ij}^m$ $\forall (i,j) \in \mathcal{A}$}
\Output{$Intflag$}
\begin{algorithmic}[1]
\STATE $count = 0$. $Intflag = 0$   \label{Op:IC1}
\FOR{$(i, j) \in \mathcal{A} \, , \, i \neq a \, , \, j \neq b$} \label{Op:IC2}
\IF{$\bigl( (p_{new,ab}^m  g_{aj}^m \geq 0.1 N_0 W x_{ij}^m)
|| (p_{ij}^m g_{ib}^m \geq 0.1 N_0 W  x_{new,ab}^m)  \bigr)$} \label{Op:IC3}
\STATE $count = count + 1$ \label{Op:IC4}
\ENDIF \label{Op:IC5}
\IF{$\bigl( x_{ia}^m + x_{ab}^m + x_{bj}^m > 0 \bigr)$} \label{Op:IC6}
\STATE $count = count + 1$ \label{Op:IC7}
\ENDIF \label{Op:IC8}
\ENDFOR \label{Op:IC9}
\IF{$count > 0$} \label{Op:IC10}
\STATE $Intflag = 1$ \label{Op:IC11}
\ENDIF \label{Op:IC12}
\RETURN   \label{Op:IC13}
\end{algorithmic}
\end{algorithm}


\subsubsection{Computational Complexity}  \label{sec:Complexity}

Global system power minimization algorithm of Algorithm~\ref{alg:GreedyMultiHop}
contains three major parts: 1) Initial routing path selection
(operation~\ref{GM:Op1}-\ref{GM:Op3}) of Algorithm~\ref{alg:GreedyMultiHop},
2) channel scheduling (operation~\ref{GM:Op9}-\ref{GM:Op18}) of Algorithm~\ref{alg:GreedyMultiHop}
along with Algorithm~\ref{alg:GreedyScheduling} and Algorithm~\ref{alg:InterferenceCheck} 
, and 3) optimal power control and routing path design (operation~\ref{GM:Op19} 
of Algorithm~\ref{alg:GreedyMultiHop}).

Initial routing path selection involves computing link weights ($O(E)$)
and shortest path ($O(E+N \log N)$) for $L$ sessions.
Therefore, the overall complexity for this part is: 
$O\bigl(L (E + N \log N) \bigr)$.

The greedy scheduling algorithm starts with a while loop. The while loop
iterates through all active links ($O(E)$). Each link calls the sub-routine
of Algorithm~\ref{alg:GreedyScheduling}, iterates through $M$ channels.
Inside each channel, the code calls the sub-routine of Algorithm~\ref{alg:InterferenceCheck}.
Algorithm~\ref{alg:InterferenceCheck} checks if the current channel
assignment interferes with other links ($O(E)$).
The global while loop of Algorithm~\ref{alg:GreedyMultiHop}
can iterate at most $O(M)$ times since each iteration will either
select a channel for a link or the loop will break.
Therefore, the greedy scheduling algorithm runs in $O(E^2 M^2)$ time.

The optimal power allocation and routing path selection
problem of operation~\ref{GM:Op19} of Algorithm~\ref{alg:GreedyMultiHop}
is a convex optimization program. Barrier method~\cite{Boyd} can solve this in 
$O(R \log(R))$ steps where $R$ is the number of inequality
constraints. From~\eqref{fig:ModifiedProblem},
we find the complexity to be $O(EM \log (EM)$. 

The greedy scheduling part dominates the overall complexity ($O(E^2 M^2)$)
of Algorithm~\ref{alg:GreedyMultiHop}. 
We term this algorithm ``GreedySysPowerMin". We compare the performance of both 
``GreedySysPowerMin" and ``BnBSysPowerMin" with that of ``TxPowerMin"
in Sec.~\ref{sec:Simulation}.

The number of channel $M$ does not denote the number of subcarriers in a multi-hop network. It
represents the number of wideband channels, e.g., $6$ MHz TV channels in white space
and $20$ MHz channels in 5.8 GHz, in the network. This ensures that 
our overall complexity ($O(E^2 M^2)$) remains reasonable for a moderately sized network.

\section{Simulation Results}  \label{sec:Simulation}

\subsection{System Power Minimization in a Single Point-to-Point Link}  \label{sec:SystemSingleHop}

\begin{figure}[t]
\centering
\includegraphics[scale=0.45]{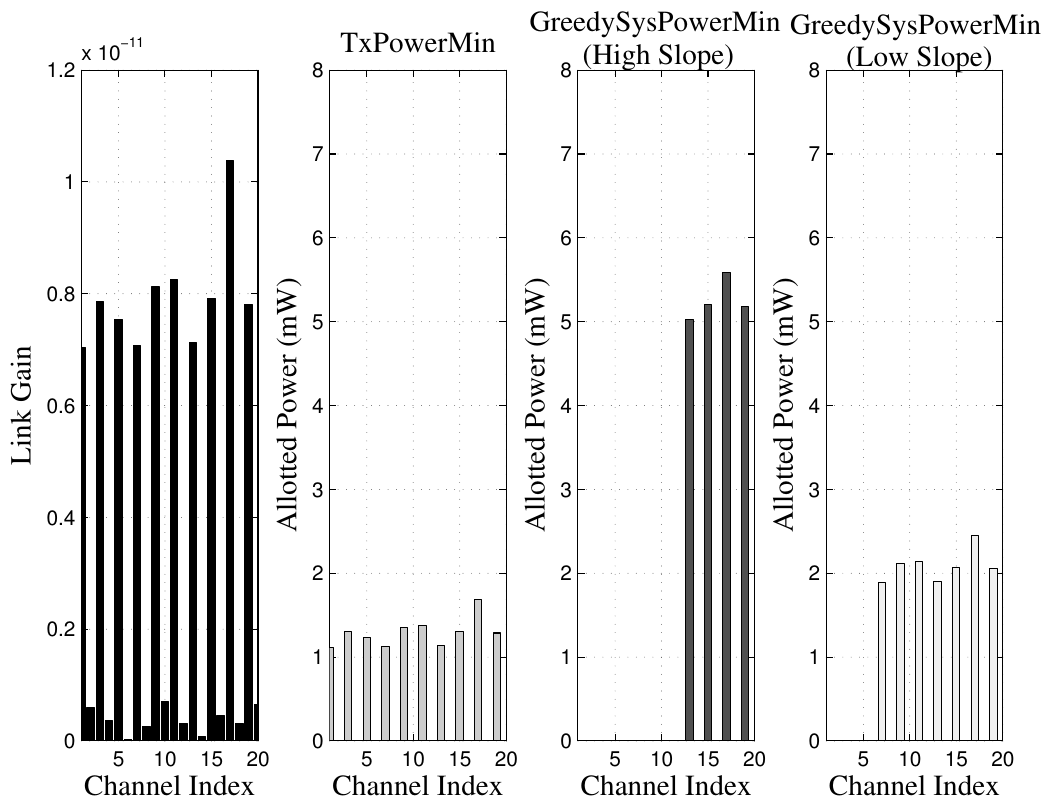}
\caption{
Comparison of `TxPowerMin' and our approach in power
allocation and scheduling  
}
\label{fig:Power_Allocation}
\end{figure}

We focus on an NC-OFDMA based single
transceiver pair. 
There is only one session in the network and 
minimum data rate requirement is $18$ Mbps. 
There are $20$ channels available for transmission. Each channel is
$3$ MHz wide. 
The left sub-plot of Fig.~\ref{fig:Power_Allocation}
shows the link gains across these channels. We designed the link
gains so that every other channel has better link gain 
than its adjacent neighbors.

The second subplot (from the left) of Fig.~\ref{fig:Power_Allocation} 
shows the power allocation
and scheduling variables of TxPowerMin approach. 
This approach minimizes transmit power subject to the rate constraint.
Similar to the concept of ``waterfilling" algorithm~\cite{Cover},
this approach spreads power across all ten ``good" channels of the network.

The third and fourth subplot (from the left) of Fig.~\ref{fig:Power_Allocation} shows the scheduling and power variables of our greedy 
algorithm (GreedySysMin). 
We use two different types of ADC and DAC models 
to investigate the influence of ADC/DAC slopes on our algorithm. 
We use the high slope ADC and DAC models -- ADC 9777 and ADS 62P4 (see
Fig.~\ref{fig:AD9777} and Fig.~\ref{fig:ADS62P4}) --  in the third
subplot of Fig.~\ref{fig:Power_Allocation} and the low slope ADC and DAC 
model -- DAC 3162 and ADS 4249 (see Fig.~\ref{fig:AD9777} 
and Fig.~\ref{fig:ADS62P4}) -- in the fourth subplot 
(the rightmost one) of Fig.~\ref{fig:Power_Allocation}.
With high slope ADC \& DAC, our algorithm focuses more on minimizing circuit power and selects only four channels. 
With low slope ADC \& DAC, 
our approach finds a trade-off between transmit 
\& circuit power and selects seven channels.

\begin{figure}[t]
\centering
\includegraphics[scale=0.5]{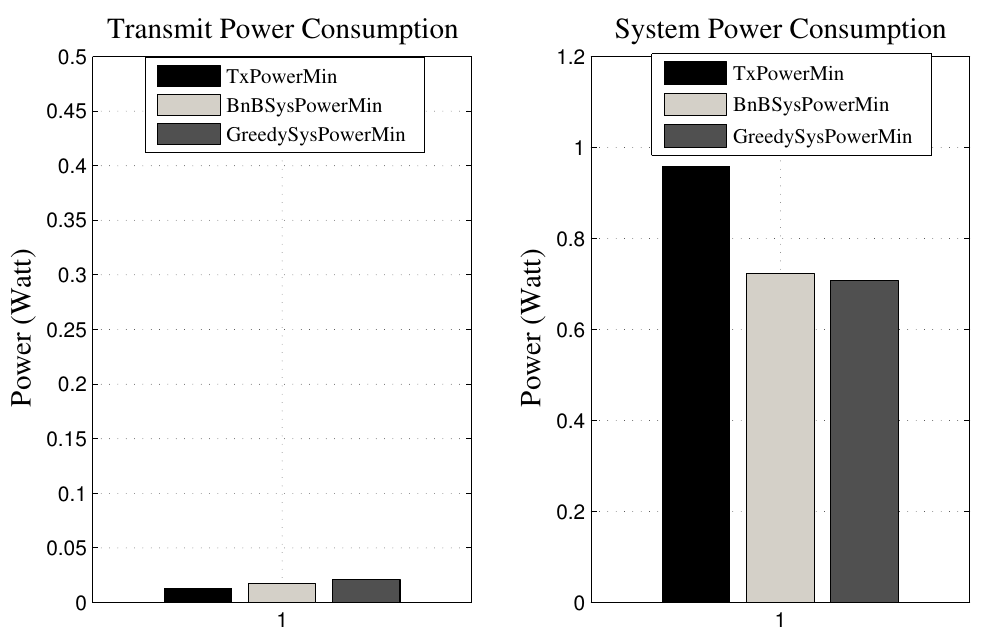}
\caption{Comparison of our algorithms with the `TxPowerMin'
approach using the high slope ADC/DAC's of Fig.~\ref{fig:AD9777} 
and Fig.~\ref{fig:ADS62P4}. 
}
\centering
\label{fig:P2PHighSlope}
\end{figure}
\begin{figure}[t]
\centering
\includegraphics[scale=0.5]{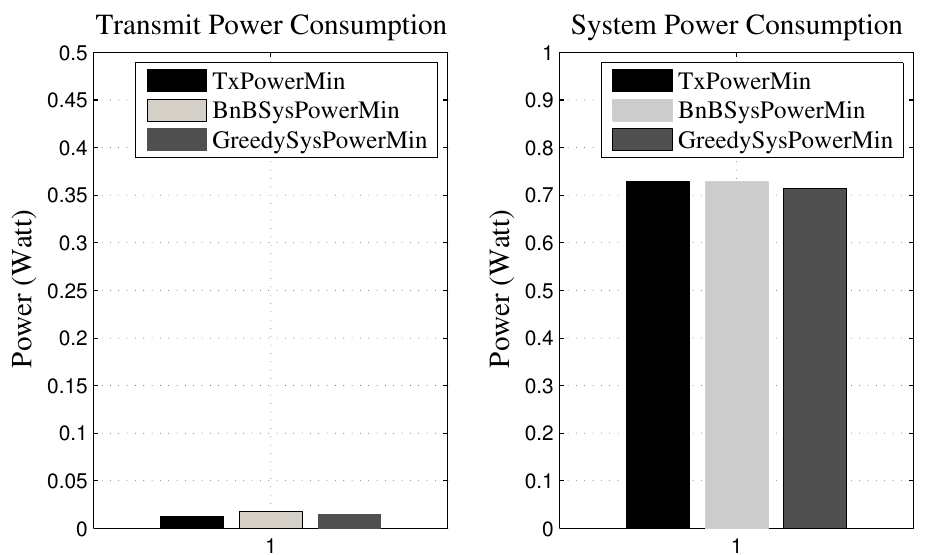}
\caption{Comparison of our algorithms with the `TxPowerMin'
approach using the low slope ADC/DAC's of Fig.~\ref{fig:AD9777} and Fig.~\ref{fig:ADS62P4}. 
}
\centering
\label{fig:P2PLowSlope}
\end{figure}

Fig.~\ref{fig:P2PHighSlope} compares the performance of our algorithm
with that of TxPowerMin
approach in high ADC/DAC slope setting. 
``TxPowerMin" approach minimizes transmit power by spreading data across all ``good" channels of the network. Both of our algorithms consume more transmit power than the ``TxPowerMin" approach since selecting a subset of available good channels 
is a sub-optimal policy in terms of transmit power.
Our algorithms consume less circuit power due to the reduced spectrum span.
Overall, our algorithms reduce system power -- summation of transmit
and circuit power -- consumption by almost $30$\%. 

Note that, the lower bound of the system power 
consumption (obtained from the mixed integer linear programming relaxation), 
was 0.63 watts in this scenario. Hence, both of
our algorithms gave feasible results with roughly 15\% optimality gap.

Fig.~\ref{fig:P2PLowSlope} compares the performance of our algorithm
with that of ``TxPowerMin"
approach in low ADC/DAC slope setting. We use the same link gain, bandwidth and traffic demands of Fig.~\ref{fig:P2PHighSlope} but we use the low power consumption ADC and DAC's  
to generate these new figures.
Our algorithm performs almost similar to the `TxPowerMin' approach here because the power consumption of these low power ADC \& DAC is negligible compared to the transmit power requirement.


\subsection{System power minimization in a multi-hop network}

\begin{figure}[t]
\centering
\includegraphics[scale=0.5]{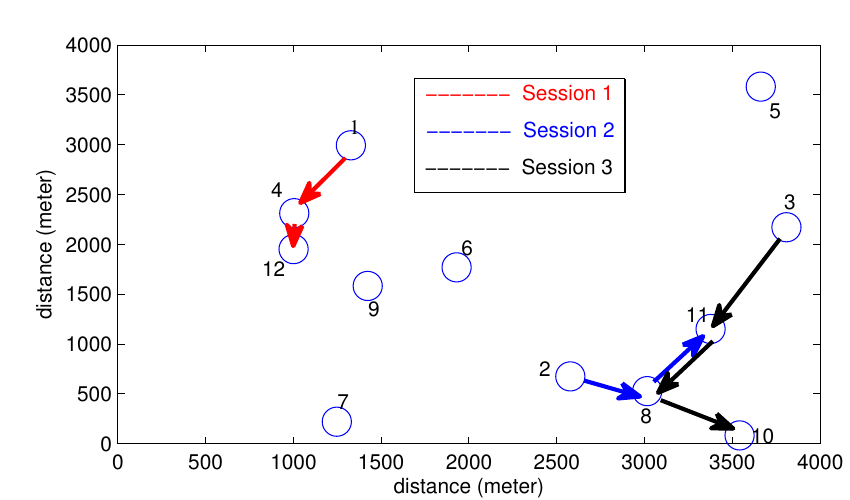}
\caption{Twelve node three session multi-hop network. }
\centering
\label{fig:Scheduling_Effect}
\end{figure}

\begin{table}[t]
\begin{center}
\begin{tabular}{|l|l|l|l|l|l|l|l|} \hline
Channel Index      & $2$ & $5$ & $6$ & $17$ & $23$ & $24$ & $47$ \\ \hline
Center Freq. (MHz) & $57$ & $79$ & $85$ & $491$ & $527$ & $533$ & $671$ \\ \hline
\end{tabular}
\end{center}
\caption{Available TV channels for fixed devices in Wichita, Kansas.}
\label{tab:ChannelList}
\end{table}

To illustrate the influence of system power minimization in a practical setting
of non-contiguous spectrum access, 
we consider an exemplary scenario of multi-hop networking among fixed devices 
in the TV white space channels of Wichita, Kansas, USA. 
We use standard spectrum databases~\cite{ShowWhiteSpace} to find the available
TV channels in Wichita, Kansas. 
Fig.~\ref{fig:Scheduling_Effect} shows the locations of nodes.
Node $1$, $2$ and $3$ transmit to node $12$, $11$ and $10$ respectively. 
Each session requires $10$ Mbps data rate.
Table~\ref{tab:ChannelList} shows the available channel indexes. Each channel is $6$ MHz wide.
We consider both large scale fading (with path loss exponent 
$3$) and small scale fading (with $12$ dB random fluctuation) in each channel. 
Maximum allowed transmission power is $4$ watts~\cite{Cyrus}.

\subsubsection{Channel Indexing Notations in Optimization Formulation}

The difference between
channel's carrier frequencies in TV bands are not always proportional to the index differences.
Channel $17$'s center frequency is ($(23-17)*6$) $36$ MHz far from
that of channel $23$ but not
($(17-6)*6$) $66$ MHz far from that of channel $6$.
The spectrum span calculation of our optimization formulations depends heavily on the coherence of channel indexing differences.
Therefore, we use an index set of
$\{9,13,14, 81, 87, 88, 111\}$ to denote the channel list of $\{2, 5, 6, 17, 23, 24, 47\}$ in the optimization formulations.
We use the original channel list to show the numerical results.\begin{table}
\begin{center}
\begin{tabular}{|l|l|l|l|l|l|} \hline
Node & Mode  & \multicolumn{2}{|c|}{TxPowerMin} & \multicolumn{2}{|c|}{BnBSysPowerMin} \\ \hline
&  & Channel & Spectrum & Channel & Spectrum \\ 
& & Index & Span  & Index & Span  \\ 
& & & (MHz) & & (MHz) \\ \hline
$1$ & Tx & $\{23,47\}$ & $150$ & $\{17,23\}$ & $42$   \\
& Rx &  $\{\emptyset\}$ & $0$ &  $\{\emptyset\}$ & $0$  \\  \hline
$2$ & Tx & $ \{17\} $ & $6$ & $\{23\}$ & $6$ \\
& Rx &  $\{\emptyset\}$ & $0$ &  $\{\emptyset\}$ & $0$ \\  \hline
$3$ & Tx &  $\{6,47\}$ & $592$ &  $\{5,6\}$ & $12$  \\
 & Rx & $\{\emptyset\}$ & $0$ &  $\{\emptyset\}$ & $0$ \\  \hline
$4$ & Tx & $ \{17\} $ & $6$ & $\{6\}$ & $6$ \\ 
 & Rx & $\{23,47\}$ & $150$ & $\{17,23\}$ & $42$   \\  \hline
$8$ & Tx & $\{2,23\}$ & $476$ & $\{2,47\}$ & $620$ \\ 
 & Rx & $ \{5,24\} $ & $460$ & $\{17,23,24\}$ & $48$ \\ \hline
$10$ & Tx & $\{\emptyset\}$ & $0$ &  $\{\emptyset\}$ & $0$  \\
 & Rx & $\{23\}$ & $6$ & $\{47\}$ & $6$ \\ \hline
 $11$ & Tx & $\{5,24\}$ & $460$ &  $\{17,24\}$ & $48$  \\
 & Rx & $\{2,6,47\}$ & $620$ & $\{2,5,6\}$ & $34$ \\ \hline
 $12$ & Tx & $\{\emptyset\}$ & $0$ &  $\{\emptyset\}$ & $0$  \\
 & Rx & $\{17\}$ & $6$ & $\{6\}$ & $6$ \\ \hline
\end{tabular}
\end{center}
\caption{Comparison between the spectrum span of `TxPowerMin' and our `BnBSysPowerMin' algorithm
in the network of Fig.~\ref{fig:Scheduling_Effect}.}
\label{tab:Scheduling_Effect_Spectrum_Span}
\end{table}

\subsubsection{Comparison of ``waterfilling" algorithm and our approach}

Here, we use the low slope ADC and DAC models of Fig.~\ref{fig:ADS62P4} and Fig.~\ref{fig:AD9777}. 
We also use Hou and Shi's algorithm~\cite{Hou} to illustrate the scheduling and power control decisions of `TxPowerMin' approach. 

\begin{figure}[t]
\centering
\includegraphics[scale=0.5]{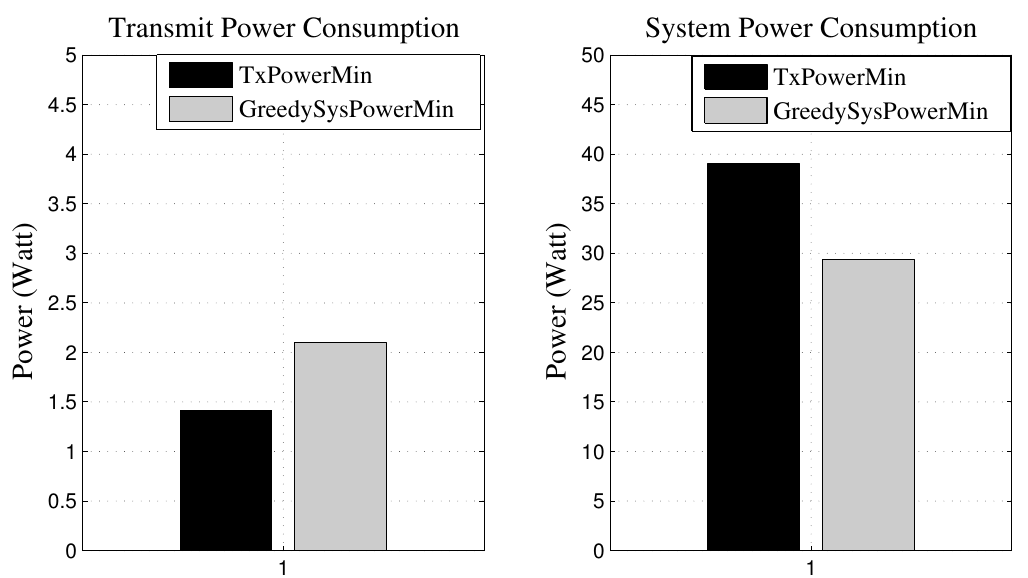}
\caption{Performance comparison of ``TxPowerMin" approach and
our algorithm in the network of Fig.~\ref{fig:Scheduling_Effect}, 
based on the low slope ADC and DAC models of Fig.~\ref{fig:ADS62P4}
and~\ref{fig:AD9777}. 
}
\label{fig:Multihop_Waterfilling}
\end{figure}
%


Table.~\ref{tab:Scheduling_Effect_Spectrum_Span} compares the channel
scheduling decisions and spectrum spans of ``TxPowerMin" approach and our algorithm. 
Although ``TxPowerMin" minimizes transmit
power by selecting channels with better quality,
it increases radio front end power by selecting channels that
are too far apart. Our approach spans 
narrow spectrum and reduces circuit power consumption. 
Fig.~\ref{fig:Multihop_Waterfilling} shows that our algorithm
reduces system power by 30\%.

\emph{The $30\%$ system power saving achieved here was obtained with 
the low slope ADC and DAC models of Fig.~\ref{fig:ADS62P4} and Fig.~\ref{fig:AD9777}
- termed as ``ultra low power ADC" and ``low power DAC" by their maker, Texas Instruments.
Had we used the high slope ADC and DAC models of Fig.~\ref{fig:ADS62P4} and Fig.~\ref{fig:AD9777},
our algorithm would have saved system power by a much higher amount in this simulation.
In the future, even if the power consumption curves of ADC and DAC become
flatter, our algorithm will reduce the system power considerably as long
as the number of available channels is sparsely located -- which is often the
case in a cognitive radio network.}

The lower bound of system power consumption is 24 watts in this scenario.
Our algorithm provides a feasible solution (29 watts) with 20\% optimality gap.

\section{Conclusion} \label{sec:Conclusion}

Wireless transmission using non-contiguous chunks
of spectrum is becoming increasingly essential.
MC-MR and NC-OFDMA are the two commercially viable choices
to access these non-contiguous spectrum chunks.
Fixed MC-MRs do not scale with increasing number of
non-contiguous spectrum chunks. 
NC-OFDMA accesses non-contiguous spectrum chunks with a 
single front end radio but
increases circuit power consumption by spanning wider spectrum.
Our approach characterized this trade-off and 
performed joint power control, channel scheduling, spectrum span selection
and routing to minimize system power consumption in an NC-OFDMA based multi-hop
network. Our algorithm showed how the slopes of ADC and DAC's power 
consumption versus sampling rate curves influence the 
scheduling decisions of a multi-hop network.
Numerical results suggested that our algorithm can save 
$30$\% system power over classical
transmission power based cross-layer algorithms 
in single front end radios. 
We developed a branch-and-bound based mixed integer linear programming 
model to optimize the cross-layer decisions of a general multi-hop network.
Furthermore, we also provided a low complexity $(O(E^2 M^2))$ 
greedy algorithm.

The optimal fragmentation results presented here have only accounted 
for the radio front end power and transmitters' emitted power. Future work will incorporate 
baseband power in the optimization formulation.

We focused on system power consumption of single front end radio in this work. 
A multi-front end programmable radio can dynamically
switch its multiple set of center frequencies and
access several spectrum chunks in each radio front end
with less spectrum span by using NC-OFDMA technology~\cite{GFDM}.
In the future, we will focus on optimal system power consumption of
multi-front end radios.

\section{Acknowledgements}

This work is supported by the Office of Naval Research under grant N00014-11-1-0132.
We thank Dr. Howard, Dr. Ackland and Dr. Samardzija for their feedback regarding 
system power consumption models.

\begin{appendices}


\section{Power Consumption of Different Blocks in Transmitter and Receiver} \label{sec:PowerConsumption}

Based on Fig.~\ref{fig:System_Block}, power consumption of transmitter and receiver 
are:
\begin{eqnarray}
p_{tc} & = & p_{dac} + p_{tfilt} + p_{mix} + p_{pa} \label{eq:TxCktPower}  \\
p_{rc} & = & p_{adc} + p_{rfilt} + p_{mix} + p_{ifa} + p_{lna}.   \label{eq:RxCktPower}
\end{eqnarray}
In the above, $p_{dac}$, $p_{mix}$, $p_{pa}$, $p_{adc}$, $p_{ifa}$ and $p_{lna}$
denote the circuit power consumption in the DAC, mixer, PA, ADC, IFA and LNA 
respectively. $p_{tfilt}$ and $p_{rfilt}$ represent the
summation of circuit powers in the filters of transmitter and receiver respectively. 

The power consumption in the mixer, LNA and IFA are 
constants with respect to the sampling rate~\cite{SystemLevelPower}. 
Baseband filter power depends on sampling rate but we assume it to be 
constant due to its low power consumption~\cite{Goldsmith}. Let
us assume, 
\begin{equation}
p_{tfilt} + p_{mix} = k_t \, , \, \, p_{rfilt} + p_{mix} + p_{ifa} + p_{lna} = k_r \, \label{eq:RxConstPower}
\end{equation}
%
%
DAC and ADC power consumptions are affine functions of
sampling rate~\cite{Goldsmith}. Hence,
\begin{equation}
p_{dac} = k_1 + k_2 fs \, , \, \, p_{adc} = k_3 + k_4 fs \, \label{eq:ADCPower}
\end{equation}
Now, using\eqref{eq:ADCPower} 
and~\eqref{eq:RxConstPower} in~\eqref{eq:TxCktPower}
and~\eqref{eq:RxCktPower}.
\begin{equation}
p_{tc} = k_1 + k_2 fs + k_t = \alpha_1 + \alpha_2 fs \, , \, \, \, p_{rc} = k_3 + k_4 fs + k_r = \beta_1 + \beta_2 f  
\end{equation}
In the above equation, $\alpha_1 = k_1 + k_t$ , $\beta_1 = k_3 + k_r$,
$\alpha_2 = k_2$ and $\beta_2 = k_4$.

We assume low power consumption at these blocks and use the following values~\cite{Goldsmith}: 
$p_{tfilt} = 5$ mW, $p_{mix} = 30.3$ mW, $p_{rfilt} = 7.5$ mW,
$p_{ifa} = 3$ mW, $p_{lna} = 20$ mW.

Due to its dependence on transmit power, we do not include
programmable amplifier's circuit power consumption term $p_{pa}$
in the overall circuit power consumption equations of~\eqref{eq:TxCktPower}
and~\eqref{eq:RxCktPower}. Instead, we couple it
with the transmit power consumption $p$ and include it
in the total power equations of~\eqref{eq:SystemPowerTx}
and~\eqref{eq:SystemPowerRx}. 



\subsection{Power consumption of programmable amplifier}


Power consumption of programmable amplifier is,  $p_{pa} = \frac{PAPR}{\eta} p$.
We assume a class-B or a higher class (C, D or E) 
amplifier with $\eta = 0.75$~\cite{Goldsmith}.

We consider OFDM to be our modulation scheme. 
The authors of~\cite{PAPRReduction3}
have related the PAPR with the number of subcarriers in OFDM systems as:
\begin{equation}
Prob\{PAPR > \gamma \} \approx 1 - exp\{-N e^{-\gamma} \sqrt{\frac{\pi}{3} \gamma} \}
\label{eq:PAPREquation}
\end{equation}
where $N$ is the number of subcarriers.



In the presence of large number
of subcarriers, the statistical distribution of the PAPR
does not remain sensitive to the increase of the number 
of subcarriers~\cite{PAPRReduction3}. During the simulations, 
we consider the worst case PAPR by assuming maximum possible
spectrum span and highest number of subcarriers.
%
We plug the value of maximum number
of possible subcarriers of TV white space in~\eqref{eq:PAPREquation}, and assume
$\gamma = 0.005$ and a $3$-$4$ dB reduction in PAPR.
Based on these calculations, we find PAPR to be around $9$ dB.

\begin{figure}[t]
\centering
\begin{subfigure}{.5\textwidth}
  \centering
  \includegraphics[width=.5\linewidth]{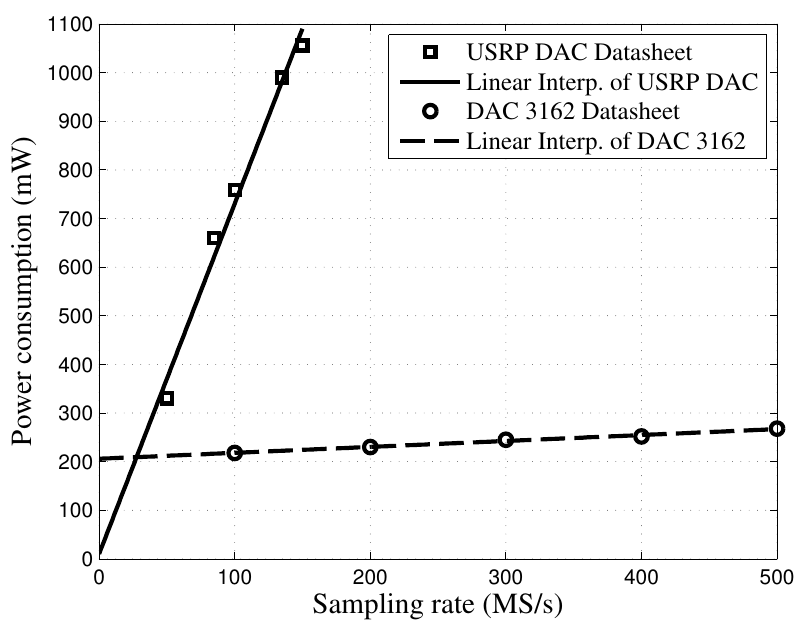}
  \caption{}
  \label{fig:AD9777}
\end{subfigure}%
\hspace{-12mm}
\begin{subfigure}{.5\textwidth}
  \centering
  \includegraphics[width=.5\linewidth]{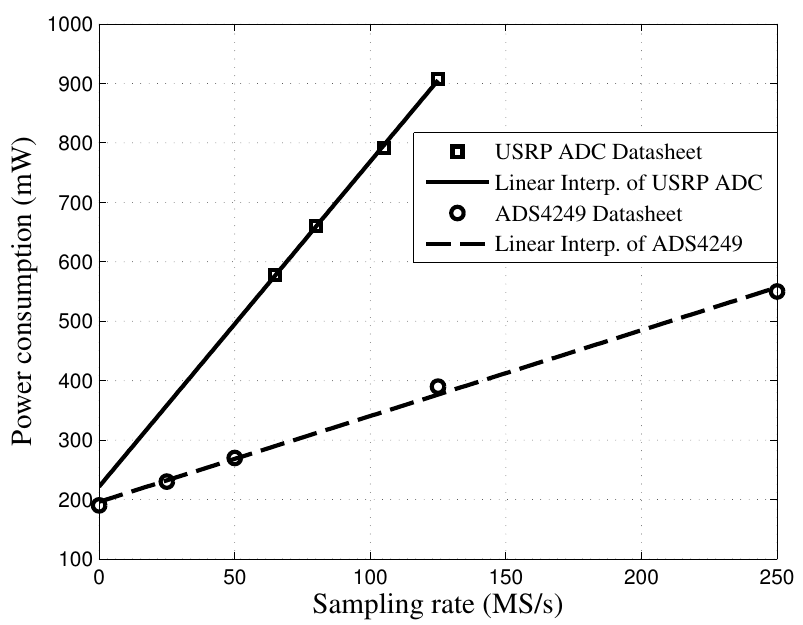}
  \caption{}
  \label{fig:ADS62P4}
\end{subfigure}
\caption{(a) Power consumption of AD 9777~\cite{USRP_DAC} (DAC of USRP radio)
and DAC 3162~\cite{LowPowerDAC}. (b) Power consumption of ADS 62P4~\cite{USRP_ADC} (ADC
of USRP radio) and ADS 4249~\cite{LowPowerADC}}
\end{figure}

%

\subsection{Power consumption of ADC and DAC}

Fig.~\ref{fig:AD9777}  plots the power consumption vs. sampling
rate curve of AD 9777~\cite{USRP_DAC} (DAC of USRP radio)
and DAC 3162~\cite{LowPowerDAC}.
Fig.~\ref{fig:ADS62P4}  plots the power consumption vs. sampling
rate curve of ADS62P4~\cite{USRP_ADC} (ADC of USRP radio)
and ADS4249~\cite{LowPowerADC}. We obtain specific values of
$k_1$, $k_2$, $k_3$ and $k_4$ from these plots.


\end{appendices}

\bibliographystyle{IEEEbib}
\bibliography{BibNCOFDMA}

\begin{thebibliography}{10}

\bibitem{NazmulCISSNew}
M.~N. Islam, A.~Sampath, A.~Maharshi, O.~Koymen, and N.~B. Mandayam,
\newblock ``Wireless backhaul node placement for small cell networks,''
\newblock in {\em Proc. {IEEE} {CISS}' 2014}, 2014, pp. 1--6,
\newblock Mar.

\bibitem{802.22}
C.~Cordeiro, K.~Challapali, D.~Birru, and S.~Shankar,
\newblock ``{IEEE} 802.22: the first worldwide wireless standard based on
  cognitive radios,''
\newblock in {\em Proc. {IEEE} {DySPAN}'2005}, Nov. 2005, p. 328–337.

\bibitem{FCCSmallCell}
``Enabling innovative small cell use in 3.5 {GHZ} band {NPRM} \& order,''
  accessed January 2016,
\newblock
  http://www.fcc.gov/document/enabling-innovative-small-cell-use-35-ghz-band-nprm-order.

\bibitem{MC-MR1}
M.~Kodialam and T.~Nandagopal,
\newblock ``Characterizing the capacity region in multi-radio multi-channel
  wireless mesh networks,''
\newblock in {\em Proc. {ACM} {MOBICOM}' 2005}, Aug. 2005, pp. 73--87.

\bibitem{MC-MR2}
P.~Kyasanur and N.~H. Vaidya,
\newblock ``Capacity of multi-channel wireless networks: impact of number of
  channels and interfaces,''
\newblock in {\em Proc. {ACM} {MOBICOM}' 2005}, Aug. 2005, pp. 43--57.

\bibitem{MCMRLimitations}
Kristoffer Fredriksson and Mattias Guhl,
\newblock ``Multi-channel, multi-radio in wireless mesh networks,''
\newblock M.S. thesis, Chalmers Institute of Technology, Goteborg, Sweden,
  2011.

\bibitem{Hou}
Y.~Shi and Y.~T. Hou,
\newblock ``Optimal power control for multi-hop software deﬁned radio
  networks,''
\newblock in {\em Proc. {IEEE} {INFOCOM}'2007}, May 2007, pp. 1694--1702.

\bibitem{Kompella}
Y.~Shi, T.~Hou, S.~Kompella, and H.~Sherali,
\newblock ``Maximizing capacity in multihop cognitive radio networks under the
  {SINR} model,''
\newblock {\em {IEEE} Transactions on Mobile Computing}, vol. 10, pp. 954--967,
  2011.

\bibitem{Hou2}
Y.~Shi and T.~Hou,
\newblock ``A distributed optimization algorithm for multi-hop cognitive radio
  networks,''
\newblock in {\em Proc. {IEEE} {INFOCOM}' 2008}, 2008, pp. 1292 -- 1300.

\bibitem{NazmulWiOpt12}
M.~N. Islam, N.~B. Mandayam, and S.~Kompella,
\newblock ``Optimal resource allocation and relay selection in bandwidth
  exchange based cooperative forwarding,''
\newblock in {\em Proc. {IEEE} {WiOpt}'2012}, May 2012, pp. 192--199.

\bibitem{Baochun}
H.~Xu and B.~Li,
\newblock ``Efficient resource allocation with flexible channel cooperation in
  {OFDMA} cognitive radio networks,''
\newblock in {\em Proc. {IEEE} {INFOCOM}'2010}, Mar. 2010, pp. 1--9.

\bibitem{UCSB2}
L.~Yang, B.~Y. Zhao, and H.~Zheng,
\newblock ``The spaces between us: Setting and maintaining boundaries in
  wireless spectrum access,''
\newblock in {\em Proc. {ACM} {MOBICOM} 2010}, Sept. 2010, pp. 37--48.

\bibitem{UCSB3}
L.~Yang, W.~Hou, L.~Cao, B.~Y. Zhao, and H.~Zheng,
\newblock ``Supporting demanding wireless applications with frequency-agile
  radios,''
\newblock in {\em Proc. 7th {USENIX} conference on Networked Systems Design and
  Implementation}, Apr. 2010, pp. 1--5.

\bibitem{Goldsmith}
S.~Cui, A.~Goldsmith, and A.~Bahai,
\newblock ``Energy-constrained modulation optimization,''
\newblock {\em {IEEE} Transactions on Wireless Communications}, vol. 4, pp.
  2349 -- 2360, SEP 2005.

\bibitem{ConverterPassion}
``{ADC} performance evolution: Walden figure-of-merit (fom),'' accessed January
  2016,
\newblock http://converterpassion.wordpress.com/2012/08/21/.

\bibitem{USRP_DAC}
``{AD9777}: 16-bit interpolating dual dac converter,'' accessed January 2016,
\newblock
  http://www.analog.com/media/en/technical-documentation/data-sheets/AD9777.pdf.

\bibitem{USRP_ADC}
``Dual channel, 14-bits, 125/105/80/65 {MSPS} {ADC} with {DDR} {LVDS/CMOS}
  outputs,'' accessed January 2016,
\newblock http://www.ti.com/lit/ds/symlink/ads62p42.pdf.

\bibitem{AD9467}
``{AD9467}: 16-bit, {200 MSPS/250 MSPS} {Analog-to-Digital} converter,''
  accessed January 2016,
\newblock
  http://www.analog.com/media/en/technical-documentation/data-sheets/AD9467.pdf.

\bibitem{Cyrus}
C.~Gerami, N.~B. Mandayam, and L.~J. Greenstein,
\newblock ``Backhauling in {TV} white spaces,''
\newblock in {\em Proc. {IEEE} {GLOBECOM}'2010}, 2010, pp. 1--6,
\newblock Dec.

\bibitem{CktPower2}
G.~Li, Z.~Xu, C.~Xiong, C.~Yang, S.~Zhang, Y.~Chen, and S.~Xu,
\newblock ``Energy-efficient wireless communications: tutorial, survey, and
  open issues,''
\newblock {\em {IEEE} Transactions on Wireless Communications}, vol. 18, pp.
  28--35, 2011.

\bibitem{Sahai}
P.~Grover, K.~A. Woyach, and A.~Sahai,
\newblock ``Towards a communication-theoretic understanding of system-level
  power consumption,''
\newblock {\em {IEEE} Journals on Selected Areas in Communications}, vol. 29,
  pp. 1744 -- 1755, SEP 2011.

\bibitem{CktPower1}
C.~Isheden and G.~P. Fettweis,
\newblock ``Energy-efficient multi-carrier link adaptation with sum
  rate-dependent circuit power,''
\newblock in {\em Proc. {IEEE} {GLOBECOMM}' 2010}, Dec. 2010, pp. 1--6.

\bibitem{Bundle}
J.~Jia and W.~Zhuang,
\newblock ``Capacity of multi-hop wireless network with frequency agile
  software defined radio,''
\newblock in {\em Proc. {IEEE} {INFOCOM} Workshop on Cognitive \& Cooperative
  Networks}, Apr. 2011, pp. 41--46.

\bibitem{GuardBand}
L.~Cao, L.~Yang, and H.~Zheng,
\newblock ``The impact of frequency-agility on dynamic spectrum sharing,''
\newblock in {\em Proc. {IEEE} {DySPAN} 2010}, Apr. 2010, pp. 1--12.

\bibitem{PNS}
Y.~P. Lin and P.~P. Vaidyanathan,
\newblock ``Periodically nonuniform sampling of bandpass signals,''
\newblock {\em {IEEE} Transactions on Circuits and Systems II}, vol. 45, pp.
  340--351, 1998.

\bibitem{SubNyquistSummary}
M.~Mishali and Y.~C. Eldar,
\newblock ``Sub-nyquist sampling,''
\newblock {\em {IEEE} Signal Processing Magazine}, vol. 28, pp. 98--124, 2011.

\bibitem{UCSB1}
L.~Yang, Z.~Zhang, W.~Hou, B.~Y. Zhao, and H.~Zheng,
\newblock ``Papyrus: A software platform for distributed dynamic spectrum
  sharing using sdrs,''
\newblock {\em {ACM} {SIGCOMM} Computer Communications Review}, vol. 41, pp.
  31--37, 2011.

\bibitem{UCSB4}
W.~Hou, L.~Yang, L.~Zhang, X.~Shan, and H.~Zheng,
\newblock ``Understanding the impact of cross-band interference,''
\newblock in {\em Proc. {ACM} Workshop on Cognitive Radio Networks}, Sept.
  2009, pp. 19--24.

\bibitem{LowPowerBaseband}
H.~C. Liu, J.~S. Min, and H.~Samueli,
\newblock ``A low-power baseband receiver {IC} for frequency-hopped spread
  spectrum communications,''
\newblock {\em {IEEE} J. Solid-State Circuits}, vol. 31, pp. 384--394, mar
  1996.

\bibitem{SystemLevelPower}
Y.~Li, B.~Bakkaloglu, and C.~Chakrabarti,
\newblock ``A system level energy model and energy-quality evaluation for
  integrated transceiver front-ends,''
\newblock {\em {IEEE} Transactions on {VLSI} Systems}, vol. 15, pp. 90--103,
  2007.

\bibitem{RLT}
H.~D. Sherali and W.~P. Adams,
\newblock {\em A Reformulation-Linearization Technique for Solving Discrete and
  Continuous Nonconvex Problems},
\newblock Kluwer Academic Publishers, Dordrecht/Boston/London, 1999.

\bibitem{CVX}
``{CVX}: Matlab software for disciplined convex programming,'' accessed January
  2016,
\newblock http://cvxr.com/cvx/.

\bibitem{Mosek}
``Mosek optimization,'' accessed January 2016,
\newblock http://www.mosek.com/.

\bibitem{Cover}
T.~M. Cover and J.~A. Thomas,
\newblock {\em Elements of Information Theory},
\newblock John Wiley and Sons, Hoboken, NJ, 2005.

\bibitem{Algorithm}
T.~H. Cormen, C.~E. Leiserson, R.~L. Rivest, and C.~Stein,
\newblock {\em Introduction to Algorithms},
\newblock The MIT Press, Cambridge,MA, 2009.

\bibitem{Boyd}
S.~Boyd and L.~Vandenberghe,
\newblock {\em Convex Optimization},
\newblock Cambridge University Press, Cambridge,MA, 1999.

\bibitem{ShowWhiteSpace}
``Show my white space,'' accessed January 2016,
\newblock http://whitespaces.spectrumbridge.com/whitespaces.

\bibitem{GFDM}
G.~Fettweis, M.~Krondorf, and S.~Bittner,
\newblock ``{GFDM} - generalized frequency division multiplexing,''
\newblock in {\em Proc. {IEEE} {VTC}'2009}, 2009, pp. 1--4.

\bibitem{PAPRReduction3}
H.~Ochiai and H.~Imai,
\newblock ``On the distribution of peak-to-average power ratio in {OFDM}
  signals,''
\newblock {\em {IEEE} Transactions on Communications}, vol. 49, pp. 282--289,
  2001.

\bibitem{LowPowerDAC}
``{Dual-Channel}, 10/12 {Bit}, 500 {MSPS} {Digital-to-Analog} converters,''
  accessed January 2016,
\newblock http://www.ti.com/lit/ds/symlink/dac3162.pdf.

\bibitem{LowPowerADC}
``{Dual-Channel}, 14-{Bit}, 250-{MSPS} {Ultralow-Power ADC},'' accessed January
  2016,
\newblock http://www.ti.com/lit/ds/symlink/ads4249.pdf.

\end{thebibliography}

\end{document}